\documentclass[12pt,a4paper,final]{iopart}

\usepackage{iopams}  
\usepackage{graphicx}
\usepackage[breaklinks=true,colorlinks=true,linkcolor=blue,urlcolor=blue,citecolor=blue]{hyperref}
\usepackage{color}

\begin{document}

\title[Direct Gyrokinetic Comparison of Pedestal Transport in JET with Carbon and ITER-Like Walls]{ Direct Gyrokinetic Comparison of Pedestal Transport in JET with Carbon and ITER-Like Walls }

\author{D. R. Hatch$^{1*}$,~M. Kotschenreuther$^{1}$,~S. M. Mahajan$^{1}$,~G. Merlo$^{1}$,~A. R. Field$^{2}$,~C. Giroud$^{2}$,~J. C. Hillesheim$^{2}$,~C. F. Maggi$^{2}$,~C. Perez von Thun$^{3}$,~C. M. Roach$^{2}$,~S. Saarelma$^{2}$,~JET contributors$^{4}$}
\address{$^1$Institute for Fusion Studies, University of Texas at Austin, Austin, Texas, 78712}
\address{$^2$CCFE, Culham Science Center, Abingdon OX14 3DB, United Kingdom}
\address{$^3$Forschungszentrum J\"ulich GmbH, Institut f\"ur Energie- und Klimaforschung--Plasmaphysik, 52425 J\"ulich, Germany}
\address{$^4$See the author list of  “X. Litaudon et al 2017 Nucl. Fusion 57 102001}
\ead{$^*$drhatch@austin.utexas.edu}

%\author{Author Two}
%\address{Address Three, Neverland}
%\ead{author.two@mail.com}

%\author[cor1]{Author Three}
%\address{Address Four, Neverland}
%\eads{\mailto{author.three@mail.com}, \mailto{author.three@gmail.com}}

\begin{abstract}

	This paper compares the gyrokinetic instabilities and transport in two representative JET pedestals, one (pulse 78697) from the JET configuration with a carbon wall (C) and another (pulse 92432) from after the installation of JET's ITER-like Wall (ILW).  The discharges were selected for a comparison of JET-ILW and JET-C discharges with good confinement at high current (3 MA, corresponding also to low $\rho_*$) and retain the distinguishing features of JET-C and JET-ILW, notably, decreased pedestal top temperature for JET-ILW.  A comparison of the profiles and heating power reveals a stark qualitative difference between the discharges: the JET-ILW pulse (92432) requires twice the heating power, at a gas rate of $1.9 \times 10^{22} e/s$, to sustain roughly half the temperature gradient of the JET-C pulse (78697), operated at zero gas rate.  This points to heat transport as a central component of the dynamics limiting the JET-ILW pedestal and reinforces the following emerging JET-ILW pedestal transport paradigm, which is proposed for further examination by both theory and experiment.  ILW conditions modify the density pedestal in ways that decrease the normalized pedestal density gradient $a/L_n$, often via an outward shift of the density pedestal.  This is attributable to some combination of direct metal wall effects and the need for increased fueling to mitigate tungsten contamination. The modification to the density profile increases $\eta=L_n/L_T$, thereby producing more robust ion temperature gradient (ITG) and electron temperature gradient driven instability.  The decreased pedestal gradients for JET-ILW (92432) also result in a strongly reduced $E \times B$ shear rate, further enhancing the ion scale turbulence.  Collectively, these effects limit the pedestal temperature and demand more heating power to achieve good pedestal performance. Our simulations, consistent with basic theoretical arguments, find higher ITG turbulence, stronger stiffness, and higher pedestal transport in the ILW plasma at lower $\rho_*$.

\end{abstract}

\vspace{2pc}
%\submitto{\PPCF}
% Comment out if separate title page not require

\section{Introduction} \label{introduction}

This work studies the change in pedestal transport that arose due to the transition from the carbon (C) to ITER-like wall (ILW) on JET~\cite{philipps_10,matthews_11}.  The results help to explain the physics processes that increase pedestal transport in JET-ILW.  The framework presented will, hopefully, be examined in more detail and built upon to optimize pedestal performance on JET-ILW and future experiments, like ITER, which may be subject to similar constraints.

The H-mode~\cite{wagner_82} pedestal drastically boosts confinement by supporting high temperatures at the plasma edge.  These high temperatures can then propagate into the core where gradients are typically strongly limited by microinstabilities.  
Nearly all prospective burning plasma tokamaks, including ITER, have been designed to exploit an edge transport barrier.  In order to achieve their aims, such devices will rely not only on the accessibility of an edge barrier but also on its robustness, measured, for example, by the attainable pedestal top temperature.  

The properties of the H-mode pedestal are governed by three complex and sensitively interconnected components: (1) MHD stability, (2) SOL and divertor conditions (closely tied to wall materials, gas fueling, strike point location, pumping, etc.), and (3) the transport (along with corresponding sources and sinks) from both neoclassical processes as well as residual micro-instabilities.  In addition to their unique parameter dependences, these components collectively exhibit intricate co-dependences that have not been well-understood.  It is, therefore, crucial to focus on parameter regimes that approximate as closely as possible reactor-relevant conditions simultaneously for all three components.

In this context, JET is at the forefront of this important aspect of fusion science due to its proximity to reactor conditions on two important fronts: its wall materials and its size (or more accurately, its dimensionless size, $\rho_* = \rho/a$, where $\rho$ is the gyro radius and $a$ is the minor radius).  ITER will achieve values of $\rho_*$ that are inaccessible in present day experiments and, due to its status as the largest operating present-day tokamak, JET can most closely approximate ITER's $\rho_*$ regime.  Moreover, since replacing C surfaces with a tungsten (W) divertor and beryllium (Be) first wall (i.e., the ILW), JET also serves as a laboratory for the effects of these wall materials.  

Since beginning operation, JET-ILW has observed a degradation of confinement~\cite{giroud_13,beurskens_13,joffrin_14,leyland_15,maggi_15,giroud_15,kim_15,nunes_16} attributable to a change in pedestal dynamics, including: a decrease in pedestal temperature; a trend toward progressively lower normalized confinement at progressively higher current and magnetic field; partial recovery of pedestal temperature and confinement with impurity seeding (in contrast to C operation where seeding degraded confinement); a demand for increased heating power to sustain high pedestal temperature; and stronger degradation of confinement than other metal wall devices~\cite{beurskens_16}.  Although much progress has been made on understanding and improving JET-ILW operation, the JET-ILW pedestal remains subject to significant constraints.

The residual pedestal transport is perhaps the least studied of the three components of the edge system.  As we will argue in this paper, it may be the missing key for understanding many recent unexplained phenomena and optimizing operation in future devices.  Transport, in combination with the relevant sources and sinks, determines the heating power necessary to achieve a given pedestal temperature; the inter-ELM evolution of pedestal density and temperature profiles, which ultimately determines the operating point at which an ELM is triggered; and the accessibility and properties of ELM-free regimes.  An understanding of pedestal transport is an indispensable component of tokamak design and operation.  

In this work we target an understanding of the transport dynamics governing the JET pedestal by studying, via linear and nonlinear global electromagnetic gyrokinetic simulations using the \textsc{Gene} code~\cite{jenko_00b,goerler_11}, a pair of representative JET pedestals---one produced in recent ILW operation and an earlier one from C operation.  The two discharges were selected to have good confinement (H98$\sim 1$) at high current (3 MA) in order to study the pedestal dynamics and transport at low $\rho_*$.  These two discharges have many comparable features but differ in the distinguishing characteristics of the transition from C to ILW, notably, significantly lower pedestal temperature for JET-ILW.  In terms of operational parameters, the discharges differ in three important ways: (1) wall materials (W/Be, C), (2) fueling rates ($1.9\times 10^{22}(e/s)$, 0), and (3) heating power (33 MW, 14.8 MW).  Consequently, the subsequent analysis is an assessment of these combined effects on transport (note that (1) and (2) are tightly coupled due to the need to gas puff in JET-ILW in order to achieve W control in steady conditions).

The overarching goal is to identify various transport processes that pertain to and distinguish the pedestal dynamics of JET-C and JET-ILW.  This work builds on previous studies of JET-ILW pedestal transport~\cite{hatch_16,hatch_17,kotschenreuther_17}, which have demonstrated several connections between gyrokinetic simulations and experimental observations. Among other things, simulations in these works showed that ion scale turbulence can arise at low values of $\rho_*$ because the velocity shear suppression is proportional to $\rho*$.  This can lead to the onset of major new transport mechanisms at low $\rho_*$.  Hence, investigations of this question on JET is crucial to projecting to future burning plasmas.  

We begin by briefly summarizing the major conclusions of this study.  Our gyrokinetic analysis of the JET-C (78697) pedestal suggests that electron temperature gradient (ETG) driven modes and microtearing modes (MTM) are the major heat transport mechanisms.  This is consistent with mounting evidence, spanning multiple machines~\cite{hatch_16,kotschenreuther_18}, that the observed fluctuations display the expected characteristics of these modes.  This combination of transport mechanisms has also been proposed and analyzed in the context of inter-ELM transport and transport at the pedestal top~\cite{dickinson_12,saarelma_13,hillesheim_15}.  Turning to the JET-ILW (92432) pedestal, we note that the distinctive features are lower temperature, shallower profile gradients, and higher $\eta = L_n/L_T$ ($L_{n,T}$ is the gradient scale length of the density, temperature, respectively).  A major cause of these changes is likely the effects of a metal wall on the density profile (both direct and indirect via the need for gas puffing)~\cite{wolfrum_17}.  Often this change in the density profile is manifest as an outward shift~\cite{stefanikova_18}, which effectively reduces the density gradient in the upper pedestal.  Higher $\eta$ translates into more robust ETG and ion temperature gradient (ITG) instabilities; in combination, demand additional heating power and limit accessible pedestal top temperatures.  Shallower gradients, particularly for the density profile, translate into weaker $E \times B$ shear, which, in combination with higher $\eta$, doubly exacerbates ITG turbulence.  Remembering that the suppression of ion scale turbulence is deemed to be a principal mechanism for the pedestal formation, the unquenched ITG turbulence considered here for JET-ILW may be unique among present day machines.  Higher values of $\eta$ (i.e. high growth rates) and lower $\rho_*$ (i.e., low shear rates) drive ITG turbulence, introduce new parameter dependences, including $\rho_*$ and impurity content, which correlate with, and may cause, certain JET-ILW trends. This framework points to several plausible routes to optimizing pedestal performance, including control of radiation, neutral penetration (linked to particle sources), and SOL and pedestal density.  
%This framework points to manipulation of SOL and pedestal density as the most promising route to optimizing pedestal performance.  A handle on these dynamics will be critical for predicting and optimizing ITER operation.   

We note that ITG instability is sensitively dependent on the ion temperature profile, which is challenging to diagnose and remains subject to considerable uncertainty.  Although we use a careful reconstruction of the ion temperature profile, the resulting predictions of ITG turbulence should be considered within the context of these uncertainties.  Nonetheless, the regime examined in this ILW discharge approaches, perhaps more closely than any other presently accessible experiment, the conditions where such turbulence would be expected to arise in the pedestal (high $\eta_i$ and low $\rho_*$).  Consequently, the prediction of ITG pedestal turbulence should be carefully examined in an iterative process between experiment, simulation, and theory.  Although many questions remain, this work, hopefully, lays the foundation for optimizing the transport and resulting confinement for JET-ILW and beyond.  

This paper is organized as follows.  Sec.~\ref{transport_paradigm} provides background information on pedestal transport and related research. Sec.~\ref{shot_description} provides descriptions of the JET pulses.  Sec.~\ref{linear} describes the linear instabilities (both local and global) identified in gyrokinetic simulations.  Sec.~\ref{nonlinear} describes the nonlinear turbulence simulations and demonstrates agreement between simulated transport levels and a careful accounting of inter-ELM power balance.  Sec.~\ref{ITG} describes the stiffness and $\rho_*$ dependences expected from ITG pedestal turbulence, and Sec.~\ref{pinch} discusses an ITG particle pinch.  Sec.~\ref{summary} provides a final discussion and summary.

\section{Pedestal Transport Paradigm} \label{transport_paradigm}

The edge of an H-mode plasma is governed by at least three interconnected processes: (1) pedestal MHD stability, (2) divertor and SOL conditions, and (3) the residual transport (along with corresponding sources and sinks) in the edge transport barrier.  Of these three, MHD stability is likely the best understood and is widely used to interpret and predict pedestal structure~\cite{snyder_09,snyder_09b}.  The pre-ELM pedestal pressure is typically observed to lie near the peeling ballooning stability boundary, and ideal MHD has been successful at describing the ultimate pressure limit of the pedestal.  There have been some notable exceptions for JET-ILW (see Refs.~\cite{maggi_15,leyland_15,bowman_18} for discussions of discrepancies with MHD limits and Refs.~\cite{aiba_NF_17,aiba_PPCF_17,giroud_18} for potential resolutions of these discrepancies).  

%Divertor and SOL conditions, which are closely tied to the wall materials, also clearly affect pedestal structure.  For example, fueling and impurity seeding are often employed to manipulate SOL conditions with various effects on the pedestal.  In particular, the pedestal density is sensitive to wall materials and SOL conditions, which can affect, among other things, the particle source in the pedestal, the separatrix density (which in turn affects pedestal density gradients), and the relative shift between the density and temperature pedestals (see, e.g. Ref.~\cite{wolfrum_17} for a summary of such effects on the density pedestal).  These details can have a strong impact on both the MHD stability and transport in the pedestal.

Divertor and SOL conditions, which are closely tied to wall materials, gas fueling, strike point location, and pumping, also affect pedestal structure.  %Fueling and impurity seeding are often employed to manipulate SOL conditions with often deleterious results for the pedestal.  
The pedestal particle source and resulting density pedestal can be strongly affected by the SOL density, fueling levels, impurity content, and wall materials (metal walls having different reflection, retention, and outgassing properties)~\cite{wolfrum_17}.  An effective outward shift of the density pedestal has also been observed in metal wall devices~\cite{dunne_17,wolfrum_17,stefanikova_18}.  On AUG, for example, this has been attributed to the so-called high-field side high-density region. These various processes can impact both the MHD stability and the pedestal transport. In the context of the ILW, many of these processes act to reduce $a/L_n$ in the pedestal (e.g., by raising the separatrix density), which has a very strong impact on the nature of microturbulence and its associated transport in the pedestal.

%These various processes can impact both the MHD stability and the pedestal transport.  In the context of the ILW, many of these processes increase the separatrix density (or otherwise effectively decrease $a/L_n$), strongly affecting drift-type pedestal instabilities.  

The residual transport in the pedestal arguably remains the least-understood component of the edge system and is the topic of this paper.  Pedestal transport determines the inter-ELM trajectory of the density and temperature profiles in relation to the heating power and particle sources.  This, in turn, determines the relative pedestal top temperature and density prior to the ELM crash.  In this section, we briefly review an emerging paradigm for pedestal transport, the mechanisms involved, and their interplay with the other elements of the edge system (MHD and SOL).  This paradigm is supported by the gyrokinetic results described below.  Many of these topics are covered at a more fundamental level in Ref.~\cite{kotschenreuther_18}.  

Plasma profile evolution is governed by conservation laws (i.e., transport equations) for the temperature, density, and momentum of each species.  These laws define the response of the profile to fluxes and sources, as illustrated, for example, in the continuity equation,    
\begin{equation}
\frac{\partial n}{\partial t} + \nabla \cdot \Gamma = S_n,
\label{continuity}
\end{equation}
where $\Gamma$ is the particle flux and $S_n$ is the particle source.  The complete picture involves such equations for the density, temperature, and momentum of electrons, main ions, and impurity species.  The focus of this work is on the most important subset of these channels, namely the density and temperature of electrons and main ions, which directly determine the confinement of energy.  %Other channels, while important, are already fairly well understood (for example, the momentum largely obeys neoclassical force balance, and impurity transport with perhaps some interesting exceptions largely follows neoclassical expectations)
The relative sources for the different channels are controlled by very different mechanisms.  The plasma temperature is maintained by outward flux from the core; particle sources include flux from NBI in the core, ionization of neutral particles from the edge and, possibly, a turbulent particle pinch~\cite{angioni_09,callen_10}; and impurities are drawn in from the edge by a neoclassical pinch.  As may be expected from such different mechanisms, the drive for the different channels also varies widely in magnitude.  A useful dimensionless metric for quantifying the drive of the various channels is the ratios of their effective diffusivities (for example $D/\chi$, where $D = \Gamma / \nabla n$, $\chi = Q / (n \nabla T)$).  The use of effective diffusivities in this manner assumes neither the absence of pinches nor the absence of nonlocal effects, but is, rather, a convenient measure of the gradient that can be supported by a given flux or source.  Alternatively, the parameter $S_n T / Q$ captures the relative drive of the two channels.            

These relative drive quantities are difficult to determine experimentally but can be estimated by interpretive edge modeling.  Such analyses have produced a consistent picture spanning multiple devices: the electron heat diffusivity far exceeds the particle diffusivity: $D/\chi_e \ll 1$~\cite{horton_05,chankin_06,callen_10}---i.e., the heat source (flux from the core) far exceeds the particle source. Although such an analysis has not been carried out for the JET discharges examined here, it has been demonstrated via EDGE2D modeling of other JET-ILW pedestals~\cite{giroud_18}.   

Inter-ELM profile evolution also offers important information about the transport behavior.  Pedestal profiles typically go through an early rebuilding period followed by a period of saturation that often makes up a substantial fraction of the inter-ELM cycle.  In nearly all cases, the temperature experiences a pre-ELM period of saturation (or near-saturation)~\cite{perez_03,saarelma_13,diallo_14,diallo_15,laggner_16,maggi_17}.  The density evolution can exhibit more variation.  On DIII-D and AUG, the density pedestal saturates early, often preceding the saturation of the temperature pedestal~\cite{diallo_15,laggner_16}.  On JET, the density often continues to evolve after the saturation of the temperature pedestal and in some cases the pedestal top density increases continuously until the ELM crash~\cite{perez_03,saarelma_13,maggi_17}.  For the JET-ILW case of interest in this paper (92432), the temperature gradient becomes limited $\sim 30\%$ into the ELM cycle, while the density gradient becomes limited $\sim 50\%$ into the ELM cycle~\cite{maggi_17}. The pressure gradient is also fixed over the last half of the ELM cycle, roughly corresponding to the period of fixed density gradient.  Both the pedestal top density and temperature increase gradually up to the ELM crash.  The continued increase in temperature is likely due to the high heating power.  Ref.~\cite{maggi_17} analyzes the inter-ELM profile evolution for this discharge as well as a wide range of other discharges spanning various heating and fueling conditions.  Although the inter-ELM evolution can be complex, the pedestal top temperature generally appears to saturate prior to pedestal top density over a substantial range of conditions.%, pointing to a heat transport mechanism as the most likely limiting factor.            

The contributors to pedestal transport must be consistent with the general picture outlined above.  For example, each prospective transport mechanism has a distinctive transport \textit{fingerprint}~\cite{kotschenreuther_18} defined by its relative impact on various transport channels. A minimum subset of pedestal transport mechanisms must include KBM (or similar MHD-like modes), ITG (or similar ion scale electrostatic modes), ETG, MTM, and neoclassical transport.  MHD-like instabilities produce equal diffusivities in all channels.  Consequently, such instabilities would fix the pressure profile at marginal stability and therefore preferentially impact the most weakly driven channel.  Since the weakest (as quantified, for example, by a parameter like $S_n T / Q$) channel is typically the density, the impact of such modes would be limited to modifying the density gradients via particle transport.  ETG and MTM affect almost solely the electron heat channel and produce negligible particle transport.  ITG (and other ion scale electrostatic modes like TEM) is more versatile, producing substantial ion (and some level of electron) heat transport and potentially producing inward, outward, or small (balanced pinch and diffusion) particle transport.  

The transport mechanisms also have distinctive dependences on other parameters, like $\rho_*$.  MTM and ETG have been found to be weakly affected by shear flow and exhibit scaling close to gyroBohm $\rho_*$ dependence~\cite{hatch_17,kotschenreuther_17}.  ITG is typically thought to be suppressed by shear flow in the pedestal but may be excited by some combination of high growth rates and low shear rates (the relevant suppression parameter is $\gamma_{ExB} / \gamma_{lin}$).  
Pedestal $E \times B$ shear rates are proportional to $\rho_*$ and sensitively dependent on profile gradients due to the dominant neoclassical force balance in the pedestal (see Sec.~\ref{ITG} for detailed discussion).  Consequently, to the extent that JET-ILW is characterized by high $\eta_i$ (high growth rates), weak density gradients (low shear), and low $\rho_*$ (low shear), it resides in precisely the parameter regime where one might expect pedestal ITG turbulence to become important.  Such was the conclusion of Refs.~\cite{hatch_17,kotschenreuther_17}, which observed a $\rho_*$ threshold roughly corresponding to JET parameters below which ITG transport is not negligible.  In agreement with basic theory, gyrokinetic simulations of pedestal ITG transport have identified the scaling $Q/Q_{GB} \propto \rho_*^{-2}$~\cite{hatch_18}, indicating a strong $\rho_*$ dependence below this $\rho_*$ threshold.  We note that, while the focus of this paper is on ITG, the dynamics may translate to any ion-scale electrostatic mode like, for example, electron drift waves or trapped electron modes (TEM). 

With this general information in hand, we are prepared to conceptually diagnose the limitations on the JET-ILW pedestal: (1) The main limitation for JET-ILW is low pedestal temperature; (2) JET-ILW demands additional heating power; (3) the pedestal top temperature generally saturates early in the ELM cycle; (4) heat diffusivities exceed particle diffusivities (not unique to JET-ILW). The culprit would seem to be a transport mechanism that preferentially produces heat transport and is amplified for JET-ILW.  The following general picture emerges and is proposed for further study on JET-ILW.  Metal walls modify the density profile via some combination of direct metal wall effects (e.g. different reflection/retention properties, or other effects on neutral penetration), the demand for gas puffing for W mitigation, or other SOL features like the high-field side high density region~\cite{wolfrum_17}.  The net effect is a weaker density gradient, which increases $\eta = L_n / L_T$~\cite{maggi_17} producing more robust ITG and ETG instabilities.  These instabilities strongly limit the pedestal temperature and demand more heating power to achieve good pedestal performance.  Crucially, both ITG and ETG preferentially produce heat flux over particle flux: ETG due to the absence of kinetic ions at electron scales, and ITG due to balanced pinch and diffusion (see Sec.~\ref{pinch} for detailed discussion).  The weak density gradient also decreases the flow shear rate, doubly affecting the ITG transport.  The improved confinement with impurity seeding may be accounted for either directly (ion dilution decreasing the turbulence) or indirectly (for example, seeding increasing pedestal particle sources or relieving the SOL conditions that weaken the density gradient).  %The stronger metal-wall constraints on JET as opposed to AUG~\cite{beurskens_16} could possibly be attributed to the $\rho_*$ dependence of ion scale turbulence (lower $\rho_*$ for the former implies weaker shear suppression), although this remains speculative and would require a dedicated study to determine in more detail.  %This is the operating hypothesis of Refs.~\cite{hatch_17} and is further supported by the present study.  %Note that KBM can be eliminated as a candidate due to its intrinsic transport fingerprint (equal diffusivities for particles and heat).  At most, it limits the density profile to keep pedestal at a KBM limit but cannot account for increased heat diffusivity.  This is not distinctive to JET-ILW and not an explanation for its distinctive properties.

At this point, two arguments against the above paradigm could be made.  First, it might be supposed that a limitation on pedestal temperature could be compensated with an increase in pedestal density in order to maintain the same pressure and MHD stability properties.  However, a limit on pedestal temperature does in fact have significant effects on the net confinement.  A trade-off of temperature for density results in increased collisionality, generally resulting in less favorable MHD stability.  For example, low pedestal temperature can feed back on MHD stability through collisionality and bootstrap current, pinning the pedestal in the unfavorable ballooning-limited regime and eliminating the advantages of shaping~\cite{maggi_15}.  Aside from MHD stability, near the Greenwald limit, it is no longer possible to compensate a temperature deficit with higher density.  Finally, with an eye toward a JET DT campaign, fusion gain is a very sensitive function of temperature (as opposed to pressure alone) in typical JET parameter regimes.  Clearly, a limit on pedestal temperature has significant implications.  A strong heat transport mechanism could also be related to slow inter-ELM recovery, discrepancies between ELM limits and peeling ballooning stability theory~\cite{maggi_15,leyland_15}, and transitions to type-III ELM regimes~\cite{bokshi_16}, although this must remain speculative as these topics have not been studied in this work.   

%A second potential argument is that an increase in pedestal transport would decrease the pedestal pressure gradient thereby broadening the pedestal and increasing the height allowed by MHD stability.  This occurs, e.g., in wide pedestal QH-mode DIII-D discharges~\cite{burrell_16}.  Such a scenario, however, is clearly not the observed behavior of JET-ILW.       

%We emphasize that this pedestal transport paradigm, though not necessarily in conflict with the MHD-centric paradigm exemplified by the EPED model, has much greater explanatory power.  EPED predicts pedestal structure by taking pedestal top density as an input and finding the pressure profile that is simultaneously at marginal stability to local KBM and global peeling-ballooning modes.  The distinguishing unexplained characteristic of the JET-ILW pedestal is the low pedestal temperature and high pedestal density.  In this context, a model that takes pedestal top density as an input is not fully predictive and may have little explanatory power.  %Moreover, EPED can say nothing about the heating power required to achieve a given pedestal temperature.  %In short, it can in principle be completely consistent with JET-ILW pressure profiles and yet offer little insight into the limitations on the JET-ILW pedestal.    %Moreover, in the context of the above discussion, KBM may be operative in the pedestal, it is clearly not responsible for the bulk of the heat transport and the resulting limitation on the pedestal temperature.    

The transport paradigm discussed here is surely not encompassing enough to account for all effects limiting JET-ILW performance.  For example, studies examining the effects of SOL conditions or pedestal MHD stability (or both in combination) do explain some aspects of the observed metal wall behavior~\cite{maggi_15,giroud_15,saarelma_15,dunne_17}.  Moreover, further work is necessary to more firmly establish the hypothesis proposed in this paper.  In particular, a detailed study of inter-ELM evolution could be conducted to understand how and when a weakened density gradient affects the transport and profile evolution.  Moreover, uncertainties in the ion temperature profile and impurity content could be mitigated with further study.  Nonetheless, the transport-oriented paradigm proposed here should become an integral part of the conceptual framework for understanding pedestal dynamics and should be actively probed by theory, computation, and experiment in future work.

In the context of these preliminary considerations, the next section describes two pedestals, JET-C (78697) and JET-ILW (92432), which will be examined within this framework.

\section{Description of JET Shots} \label{shot_description}

Pedestal temperatures and normalized confinement times (relative to IPB98y2) are typically reduced in ILW plasmas relative to C, particularly at higher current and field~\cite{nunes_16}, though the degradation of confinement is not universal in ILW plasmas and it can also be mitigated.
JET pulse number 92432 was selected for this study since it represents a breakthrough in extending good confinement ($H_{98}=1$) to higher current (3 MA) and magnetic field (2.8 T) than was previously possible~\cite{kim_18}.  This is attributable to low fueling (lowest allowable for W control, albeit still substantial), high power to compensate for reduced pedestal temperature, and strike points located in the divertor corner for optimal pumping and density control~\cite{kim_18}.  An increased ratio of ion to electron temperature in the core also contributes to the improved confinement.     

\Table{\label{tab1} Summary of important parameters for JET pulses 78697 (C) and 92432 (ILW).  $I_p$ is the plasma current, $B_T$ is the toroidal magnetic field, $q_{95}$ is the safety factor at $95\%$ of the minor radius (in terms of normalized poloidal magnetic flux), $\delta$ is triangularity, $P_h$ is the total heating power, $P_i$ is the inter-ELM power loss~\cite{field_18}, Gas is the fueling rate, $\beta_N$ is normalized plasma pressure, $T_{e,p}$ is the pedestal top electron temperature, and $n_{e,p}$ is the pedestal top electron density.  }
\br
\centering
Pulse & $I_p(MA)$ & $B_T(T)$ & $q_{95}$ & $\delta$ & $P_{h} (MW)$ & $P_{i} (MW)$ & Gas(e/s) & $\beta_N$ & $T_{e,p}(keV)$ & $n_{e,p} (10^{19}m^{-3})$  \\
\mr
92432 & 3.0  & 2.8 & 3.0 & 0.2 & 33.0 & 11.6 & $1.9\times10^{22}$ &2.15 & 1.1 & 5.86   \\
78697 & 3.0  & 2.4 & 2.6 & 0.24 & 14.8 & 5.7 & 0.0 &1.8 & 1.68 & 4.19   \\
\br
\end{tabular}
\end{indented}
\end{table}

\begin{figure}[htb!]
 \centering
 \includegraphics{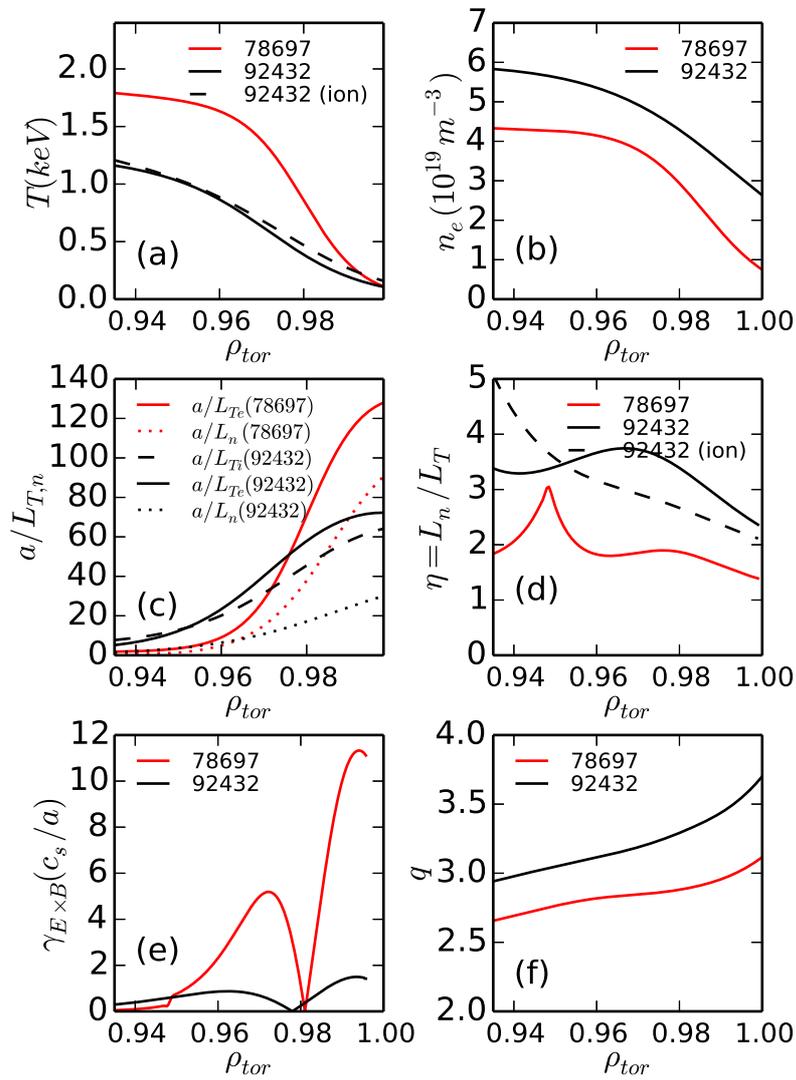}
 \caption{\label{profiles}  Profiles for JET-ILW (black) and JET-C (red) of temperature (a), density (b), $a/L_n$ and $a/L_T$ (c), $\eta = L_n/L_T$ (d), $E \times B$ shear rates (e), and safety factor $q$ (f).  Notably, JET-ILW has weaker gradients, higher $\eta$, and lower shear rates. }
\end{figure}

JET-C pulse number 78697 was also selected as a high current (3 MA) discharge with good confinfment, and matches many of the relevant parameters of 92432 while retaining a higher pedestal temperature.  It is important to keep in mind that, although these shots are representative in many ways, they clearly provide only a snapshot of the variation within and between each class of discharges.  Important parameters are listed in Table~\ref{tab1}, which demonstrates the three main operational differences in the discharges, (1) wall materials, (2) fueling rate, and (3) heating power.  Fig.~\ref{profiles} shows the pedestal profiles of many of the important quantities.  Notably, although the pedestal pressure is comparable between the two discharges, the total stored energy is higher in the JET-ILW pulse (8 MJ as compared to 6 MJ for the JET-C pulse) due to improved core confinement~\cite{kim_18}. Temperature and density profiles are shown in Fig.~\ref{profiles} (a) and (b).  The main ion (deuterium) temperature is available for 92432 and shown in Fig.~\ref{profiles} (a).  The edge ion temperature is measured from the coupled CVI and NeX CX spectrum (the Ne comes from diagnostic puffs used to improve the quality of the CX data, given the very low C concentration in JET-ILW).
The main ion species is then assumed to be thermally equilibrated with the impurities through collisional relaxation.  This measurement is very challenging in the pedestal and subject to substantially uncertainty.  For example, recent work on DIII-D exploiting a direct measurement of the main ions~\cite{haskey_RSI_18} has shown that the main ion temperature can deviate substantially from the impurity temperature in the pedestal sometimes resulting in a steeper gradient~\cite{haskey_PPCF_18}. Ion temperature measurements are not available for 78697, for which we assume $T_i=T_e$ (this effectively places an upper bound on the role of ITG in 78697, which is found to be negligible).  Figs.~\ref{profiles} (c), (d), and (e) show several important quantities for the microinstabilities.  The normalized density and temperature gradients (c) are much larger for JET-C (78697).  This translates into much higher $E\times B$ shear rates (e) for JET-C (78697) due to the dominant neoclassical force balance between $E_r$ and the profile gradients (see Sec.~\ref{ITG} for a detailed description of the calculation of shear rates).  Shear rates can also be affected by the toroidal rotation, which is reduced by gas puffing.  Further work is needed to determine what combination of gas puffing and direct metal wall effects contribute to establishing the shear rate.  Notably, $\eta$ (d) is significantly larger for ILW than C due to a large decrease in density gradient on ILW.  The $q$ profile is shown in Fig.~\ref{profiles} (f) and is less steep (i.e., lower magnetic shear) for JET-C (78697) due to the lower collisionality and higher bootstrap current.  The inter-ELM profile evolution is described in Sec.~\ref{transport_paradigm} and described more generally in Ref.~\cite{maggi_17}.  Fig.~\ref{profiles} will be referred to throughout the paper as the impact of these profiles on the relevant transport mechanisms is discussed.  

From the information in Table~\ref{tab1} and Fig.~\ref{profiles}, one can quickly glean a striking difference in the transport between the two cases: the JET-ILW (92432) pedestal requires twice the heating power to achieve approximately half the temperature gradient and similar pedestal top pressure.  To quantify this difference, the pedestal heat diffusivities are $\chi = Q/(n \nabla T) = 0.48 m^2/s$ and $\chi = 0.12 m^2/s$ for 92432 and 78697 respectively (quantities are taken in the middle of the steep gradient region of the pedestal). 
%(the difference is even starker for the respective gyroBohm normalized quantities $\chi/\chi_{GB} = 4.14$ and $\chi/\chi_{GB} = 0.29$).  
In these expressions, $Q$ is the heat flux, $n$ the electron density, $T$ the electron temperature.
%, and $\chi_{GB} = a c_s \rho_*^2$, where $c_s=(T_e/m_i)^{1/2}$ is the sound gyroradius, $a$ is the minor radius, and $\rho_*$ is the ratio of the sound gyroradius to the minor radius defined locally in the pedestal .  
From this quick comparison, one may suspect that a complete explanation of JET-ILW pedestal dynamics must include an understanding of these substantial differences in pedestal transport, which is the goal of this work. As described in the previous section, this paper proposes an emerging framework for the causes and manifestations of this change in transport.  A more detailed understanding of the divertor, SOL, neutrals, materials considerations (etc) that cause the change in transport is equally important albeit outside the scope of this work.

\section{Linear Gyrokinetic Analysis} \label{linear}

We begin our gyrokinetic investigation with an examination of the underlying linear instabilities.  Nonlinear simulations estimating the impact of these instabilities on the transport are described in Sec.~\ref{nonlinear}. 

\subsection{Local} \label{local}

The local flux tube approximation takes parameters at a single radial position, assumes constant gradients, and does not account for variation in background quantities.  This can provide qualitative information but can also be misleading in systems where background quantities vary over relatively short radial scales.  Critically, the local treatment knows nothing about the radial width over which a physics regime is operative---i.e., the effective $\rho_*$~\cite{mcmillan_10}, which is determined by equilibrium variation over the steep gradient region.  %A local approximation is also ignorant of the alignment of rational surfaces (which can be sparsely populated at low n) with background gradients.  Such information, is indispensable for the treatment of many important pedestal instabilities.  
Despite these shortcomings, the local analysis provides valuable insights, makes connections with previous literature, and demonstrates qualitative differences between ILW and C.  A more comprehensive global treatment is discussed in the next subsection.

\begin{figure}[htb!]
 \centering
 \includegraphics{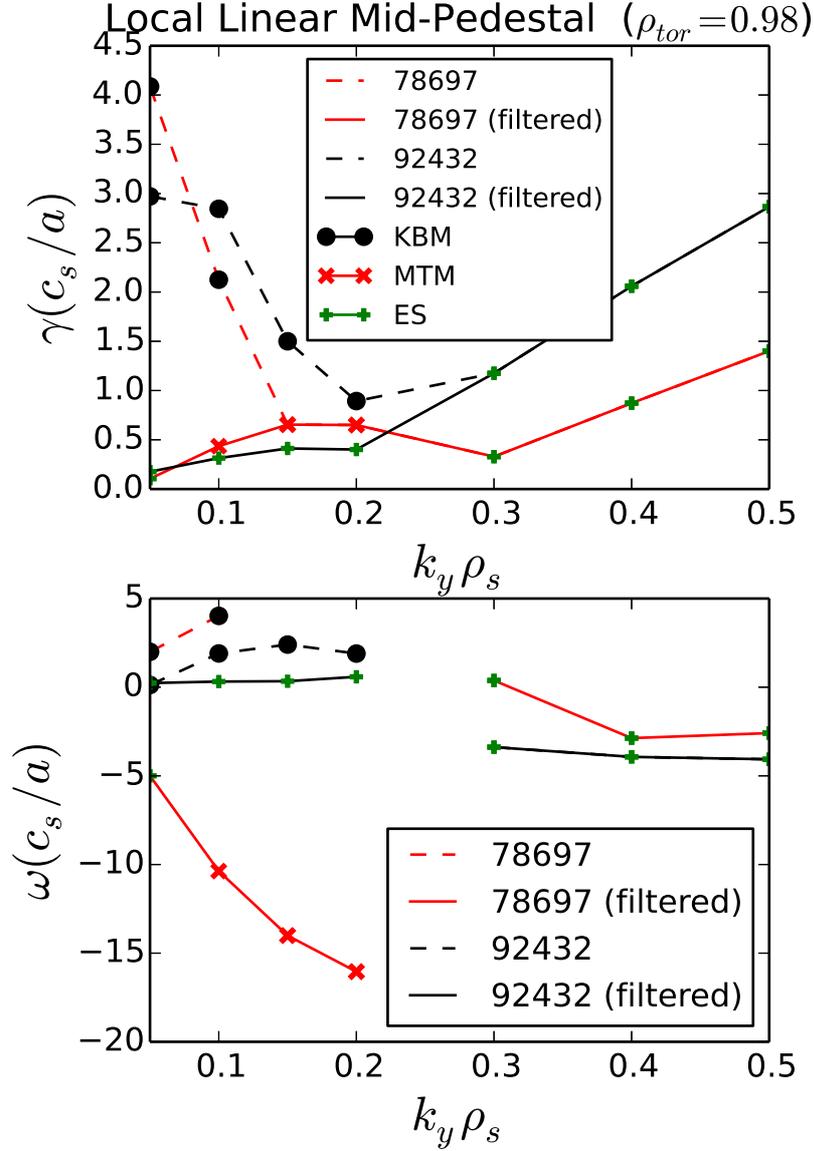}
 \caption{\label{local_linear_x98}  Growth rates (top) and frequencies (bottom) from local linear simulations in the steep gradient region for JET-ILW (92432) (black lines) and JET-C (78697) (red lines).  Note that negative frequencies correspond to the electron diamagnetic direction.  KBM, MTM, and electrostatic instabilities are observed.  The KBM have very broad radial mode structures that are inconsistent with the width of the pedestal.  The dashed lines, which find much closer agreement with global simulations, denote the instabilities that remain after filtering unphysically large radial modes.   }
\end{figure}

%\begin{figure}[htb!]
% \centering
% \includegraphics{ETG_gammas.ps}
% \caption{\label{ETG_gammas}  Electron scale growth rates for JET-ILW and JET-C.  }
%\end{figure}

Fig.~\ref{local_linear_x98} shows growth rates and frequencies for JET-C (78697) and JET-ILW (92432) from local linear simulations in the middle of the steep gradient region of the pedestal ($\rho_{tor} = 0.98$).  For both cases, KBM (black symbols) is observed at low $k_y$ and electrostatic modes are observed at high $k_y$.  For JET-C (78697), MTM is also observed at low $k_y$.  The electrostatic instabilities at higher $k_y$ have much higher growth rates for JET-ILW (92432), a trend that extends to ETG instabilities at electron scales (not shown).  This local analysis involved scans of both binormal wavenumber $k_y$ and ballooning angle ($\theta_0$).  The figure shows the maximum growth rate at each $k_y$.  In the pedestal, the scan of ballooning angle is critical.  Modes that are driven by toroidal resonances (i.e. resonances with the curvature and $\nabla B$ drifts), which is standard in the core plasma, peak at $\theta_0$.  In contrast, due to the steep gradients in the pedestal, $\omega_*$ greatly exceeds the magnetic drift frequencies and slab resonances (with $v_{||}k_{||}$) can occur.  This can produce modes with growth rates that have only weak dependence on, or even increase with, $\theta_0=0$.  The MTM shown in Fig.~\ref{local_linear_x98}, for example, are only the most unstable modes at $\theta_0>0$.  Refs.~\cite{hatch_16,kotschenreuther_17,kotschenreuther_18} provide more discussion on this topic along with plots of the $\theta_0$ dependence of growth rates. 

The electrostatic modes and the MTMs will be discussed in more detail in the context of global simulations below (Sec.~\ref{global}).  KBM, however, is not identified in the global simulations.  We focus now on its the potential role in these discharges.  

A local description of strongly ballooning modes (i.e., modes peaked strongly at the outboard midplane) can be highly suspect in the pedestal, since such modes can have mode structures whose radial extent far exceeds the width of the pedestal.  This consideration can be quantified by comparing the effective radial extent of the mode, as quantified by an eigenmode-weighted radial wavelength $1/ \langle k_x \rangle$, with the width of the pedestal as encompassed in the following metric~\cite{kotschenreuther_18}:    
\begin{equation}
\Theta = w \langle k_x \rangle,
\label{deltakx}
\end{equation}
where $w$ is the width of the pedestal and $\langle k_x \rangle$ can be calculated from the linear eigenmode structure as 
\begin{equation}
\langle k_x \rangle = \frac{\sum_{k_x} |k_x| |\phi|^2_{k_x}}{\sum_{k_x} |\phi|^2_{k_x}},
\label{k_x_sum}
\end{equation}
so that a mode that is strongly peaked at the outboard midplane ($k_x = 0$) has a very low value of $\Theta$.  Flux tube simulations with \textsc{Gene} include several $k_x$ modes (each having a parallel domain spanning $z=(-\pi,\pi)$) connected with the flux tube parallel boundary condition~\cite{beer_95}.  The ballooning angle $\theta_0 = \frac{k_{x,center}}{\hat{s} k_y}$ (i.e., the central $k_x$ value) can be chosen for each linear flux tube simulation and the the remaining $k_x$ values, $\Delta k_x = 2 \pi \hat{s} k_y$, are defined by the parallel boundary condition.  The connected $k_x$ modes constitute the extended ballooning structure of the mode so that the sum over $k_x$ defined in Eq.~\ref{k_x_sum} weights each $z=(-\pi,\pi)$ segment with its corresponding value of $k_x$.  The criterion $\Theta > 2$ indicates whether a mode predicted by local theory plausibly fits within the radial range over which it is driven.  If this criterion were to hold, it would indicate whether the local mode could in principle be physical (i.e. whether the local mode can be captured by a higher fidelity global treatment of the system).  When this criterion is applied to the modes shown in Fig.~\ref{local_linear_x98}, it eliminates the KBM; such modes are, in fact, not observed in the global simulations described below.  

Several studies examining KBM stability in the pedestal yield widely varying conclusions: KBM is unstable, albeit close to marginal~\cite{dickinson_12}; KBM is locally stable in most of the pedestal owing to the bootstrap current giving access to 2nd stability~\cite{saarelma_13}; KBM is unstable~\cite{diallo_14}; KBM is unstable but subdominant to electrostatic modes~\cite{wang_12}; KBM is stable but near the instability threshold when $\beta$ is artificially increased~\cite{hatch_15}; KBM is stable and remains stable when self-consistently varying the equilibrium consistent with increasing pressure gradients (i.e., KBM is in a second stability regime)~\cite{canik_13}; KBM is in a second stability regime locally, which disappears in a global treatment~\cite{saarelma_17}; KBM is in second stability in both local and global analysis when self consistently increasing pressure gradients with equilibrium~\cite{hatch_16}.  

In the context of this study we can make the following statements about pedestal modes: (1) KBM is identified in local simulations but has mode structures inconsistent with the narrow width of the pedestal and is not found in our global treatment, and (2) In the magnetic fluctuation data available for JET, one can identify fluctuations that appear to correspond closely with MTM and eliminate KBM on the basis of simple frequency comparisons.  In spite of these considerations, KBM (or some similar MHD activity) cannot be ruled out because it may be only marginally unstable and due to several other considerations, including uncertainties in input data, the limited fluctuation diagnostics available, and limitations of pedestal gyrokinetics for low-n MHD.  Regarding the latter, a rigorous MHD analysis would include the vacuum solution and the kink drive term, both of which are planned for future work.  Perhaps the best evidence of some sort of KBM-like MHD behavior is its apparent utility (or the utility of its proxy relation between $\beta$ and pedestal width) in the EPED model and the observation that inter-ELM profile evolution often allows separate density and temperature changes constrained by a fixed pressure profile (see, for example, Refs.~\cite{hatch_15, maggi_17}).  In any case, the transport fingerprint of KBM~\cite{kotschenreuther_18} eliminates it as a major heat transport mechanism due to its large ratio of $D_e/\chi_e$.  It can therefore be ruled out as the mechanism limiting the pedestal temperature in JET-ILW and is not of primary interest in this work.  

%Despite its limitations, the local linear analysis discussed here already illustrates some basic observations that outline the major difference between our JET-ILW and JET-C pedestals, which will be reinforced in the global discussion below: JET-ILW exhibits more robust electrostatic instabilities and JET-C exhibits MTM instability at low $k_y$.    

\subsection{Global} \label{global}

\begin{figure}[htb!]
 \centering
 \includegraphics{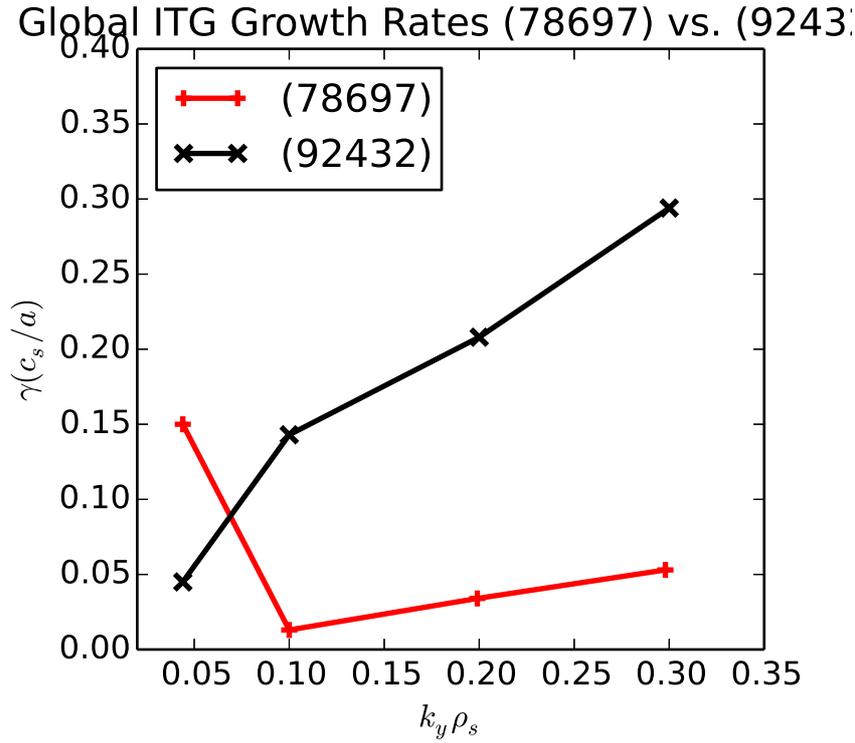}
 \caption{\label{gamma_ky}  Growth rates from global simulations for JET-ILW (92432) (black line) and JET-C (78697) (red line).  The instabilities are ITG modes, with the exception of the lowest $k_y$ JET-C (78697) mode, which is an MTM. }
\end{figure}

Having briefly surveyed the local linear gyrokinetic stability picture, we turn now to a global treatment, which accounts for the strong profile variation over the narrow pedestal.  Background rotation (described in more detail in Sec.~\ref{ITG}) is included in these global linear simulations (in contrast with the local simulations in which background shear produces Floquet modes).  Global $k_y$ scans comparing ILW (92432) and C (78697) are shown in Fig.~\ref{gamma_ky} where green symbols denote electrostatic (ITG) modes and the red x denotes a MTM observed at low $k_y$ in the JET-C (78697) pedestal.  This MTM occurs at toroidal mode number $n=8$, at which a magnetic fluctuation can be observed in magnetic spectrograms for this discharge (this will be described in detail in a future paper).  The large difference in $\eta$ for JET-C (78697) and JET-ILW (92432) is reflected in a corresponding difference in ITG growth rates.  This is further enhanced by the much larger JET-C (78697) shear rate (see Fig.~\ref{profiles} (e)), suggesting that the ion scale ITG transport for JET-ILW (92432) is expected to be much higher than that of JET-C (78697).  % The large discrepancy in ITG growth rates is particularly important in this context; not only do the smaller JET-C growth rates imply smaller drive but the shear suppression parameter $\gamma_{E\times B} / \gamma_{lin}$ is also much higher for JET-C due to its much higher shear rate (see Fig.~\ref{profiles} (e). 

\begin{figure}[htb!]
 \centering
 \includegraphics{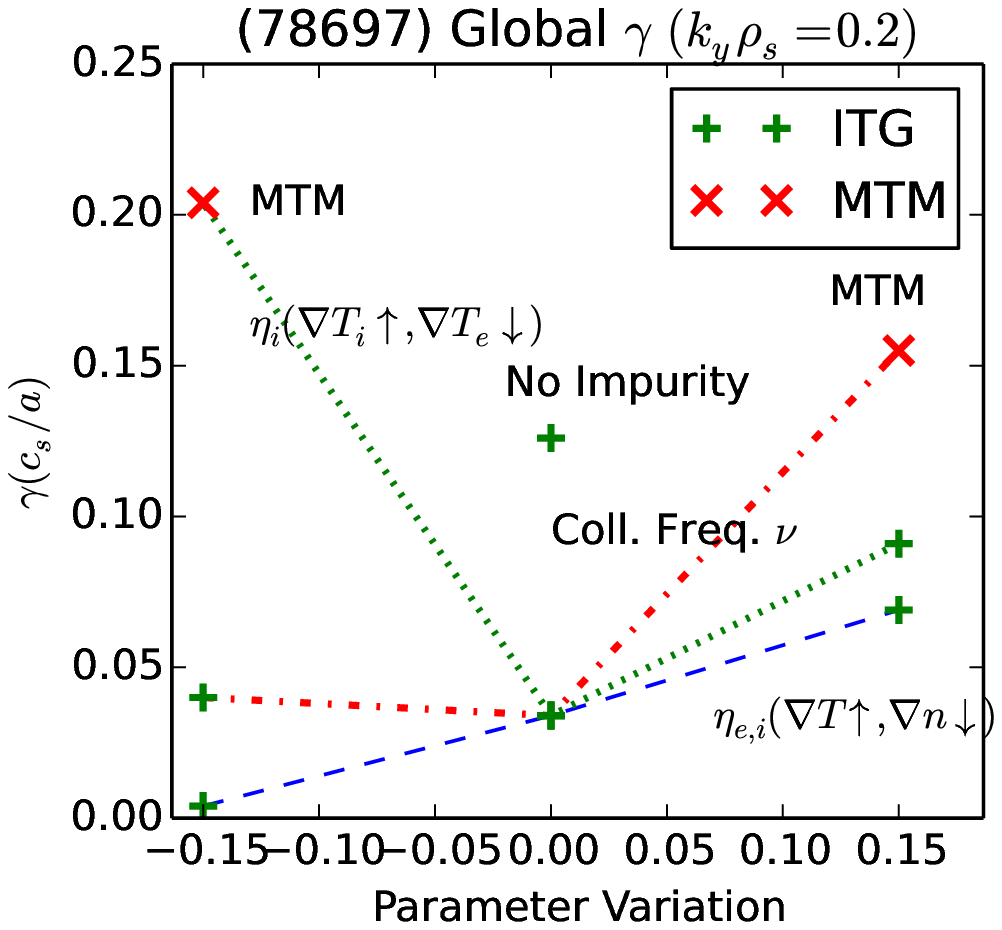}
 \caption{\label{sensitivities7}  Parameter sensitivities from global simulations at $k_y \rho_s=0.2$ for JET-C (78697).  The x-axis denotes the fractional change ($\pm15\%$) in various parameters (annotated in the figure).  All parameter variations are constructed to keep the pressure fixed (e.g., the annotation $\eta_i (\nabla T_i \uparrow , \nabla T_e \downarrow)$ denotes an increase in $\eta_i$ that keeps pressure fixed by increasing the ion temperature gradient and decreasing the electron temperature gradient). The instability found with nominal inputs has the parameter dependences of an ITG mode.  An MTM becomes dominant with minor increases in collision frequency and electron temperature gradient.  Sensitivity to impurities is demonstrated by the large increase in growth rate (labeled No Impurity) observed when C is not included. }
\end{figure}

\begin{figure}[htb!]
 \centering
 \includegraphics{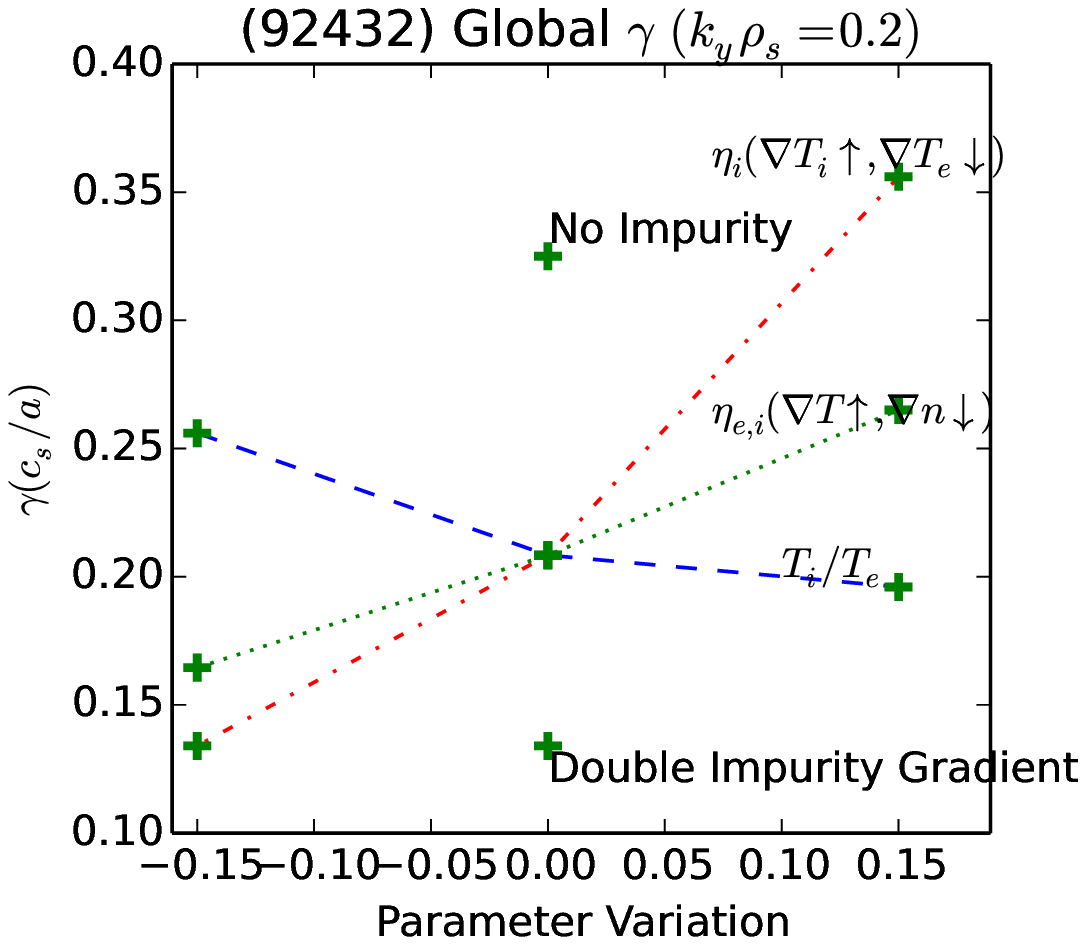}
 \caption{\label{sensitivities9} Parameter sensitivities from global simulations at $k_y \rho_s=0.2$ for JET-ILW (92432).  The x-axis denotes the fractional change ($\pm15\%$) in various parameters (annotated in the figure).  All parameter variations are constructed to keep the pressure fixed (e.g., the annotation $\eta_i (\nabla T_i \uparrow , \nabla T_e \downarrow)$ denotes an increase in $\eta_i$ that keeps pressure fixed by increasing the ion temperature gradient and decreasing the electron temperature gradient). The instability clearly exhibits the expected parameter dependences of an ITG mode. Sensitivity to impurities is demonstrated by the large increase in growth rate (labeled No Impurity) observed when Be is not included.}
\end{figure}

The nature of the instabilities is explored in Figs.~\ref{sensitivities7},\ref{sensitivities9}, which show several parameter variations for $k_y \rho_s = 0.2$ whose dependences are consistent with ITG and MTM.  All parameter variations are constructed to hold pressure fixed (e.g. increases in temperature gradient are compensated by decreases in density gradient, etc.) and thus avoid decoupling profile effects from equilibrium effects in an inconsistent manner.  Growth rates of the electrostatic modes increase with ion temperature gradients and temperature ratio $T_e/T_i$ and are insensitive to collision frequency.  MTMs arises in these scans for the JET-C (78697) pedestal as the collision frequency or electron temperature gradient is increased.  As seen in Fig.~\ref{gamma_ky}, the JET-C (78697) pedestal lies above the MTM threshold for other $k_y$ suggesting that MTM play an important role in this JET-C (78697) pedestal.  A detailed analysis of the MTM in 78697 will be presented in a future paper.

The ILW simulations include a dynamic Be impurity at the level of $Z_{eff}=1.8$ and the C simulations include a dynamic C impurity with $Z_{eff}=2.35$.  The $Z_{eff}$ estimate is derived from a line averaged measurement from visible Brehmstrahlung, which provides no profile information (constant $Z_{eff}$ is thus assumed).  The impurity content is subject to considerable uncertainties.  For example, for JET-ILW (92432) the diagnosed value of $Z_{eff}$ is produced by some mixture of impurities, including Be and Ni, but the precise densities are unknown. The instabilities can be highly sensitive to the impurity content; as shown in Fig.~\ref{sensitivities9} the ITG growth rate increases by $\sim 65 \%$ when ion dilution is neglected (i.e. no impurity is included, or alternatively assuming the $Z_{eff}$ is attributable mostly to high Z Ni), and decreases by $\sim 40 \%$ when the impurity gradient is doubled.  Likewise, the ITG growth rate in JET-C (78697) increases by a factor of four when C is neglected, as seen in Fig.~\ref{sensitivities7}.  This reinforces similar observations in Ref.~\cite{hatch_17,kotschenreuther_17} and is consistent with the beneficial effects of impurity seeding in the ILW pedestal.  

\begin{figure}[htb!]
 \centering
 \includegraphics{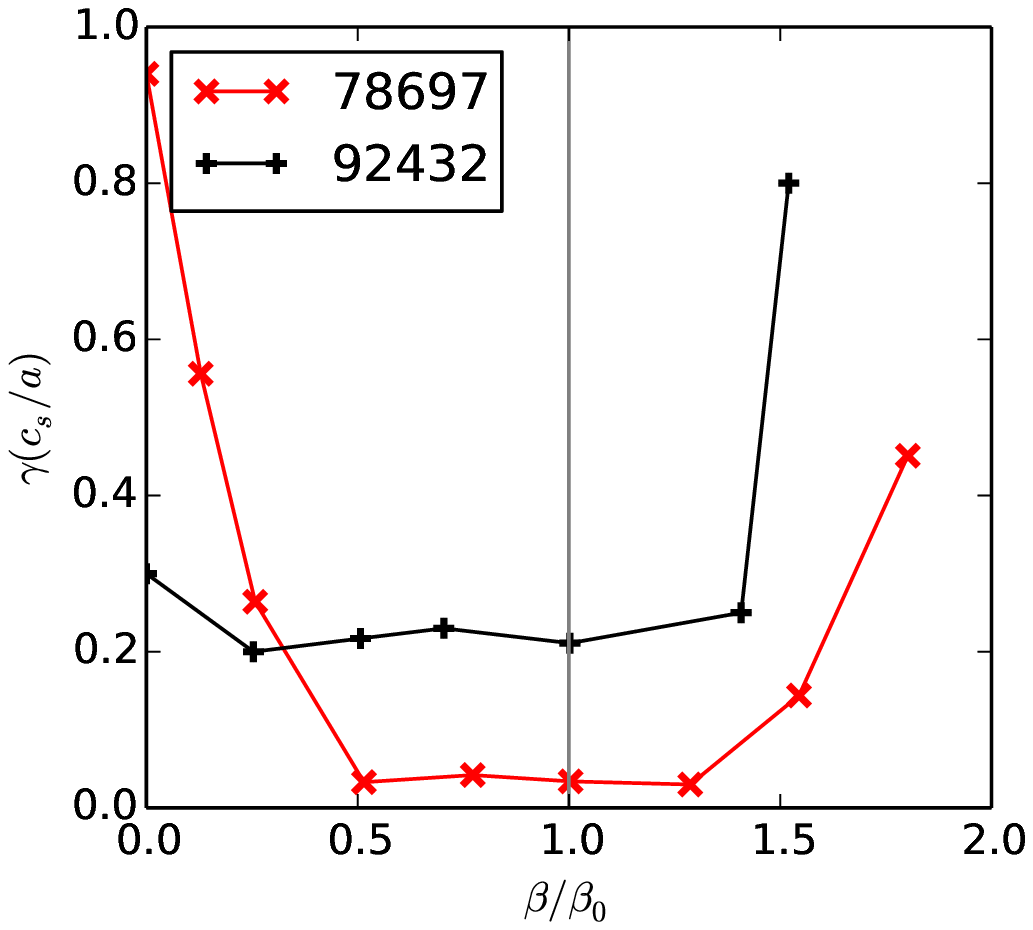}
 \caption{\label{beta_scan}  Growth rates from global simulations scanning $\beta$ (not accounting for self-consistent variation of the equilibrium) at $k_y \rho_s = 0.2$ for JET-ILW (92432) (black) and JET-C (78697) (red) demonstrating the insensitivity of the slab ITG modes to $\beta$.  However, some level of electromagnetic effects are important to suppress a highly unstable electrostatic mode that is manifest at low $\beta$.  The profiles lie $\sim 40 \%$ below a KBM limit, indicated by the sharp rise at higher $\beta$.}
\end{figure}

Notably, KBM is not observed in the global simulations, consistent with the criterion described above for local simulations.  Scans of $\beta$ were undertaken in order to probe proximity to the KBM limit and more generally to determine the sensitivity of the various modes to electromagnetic effects.  As shown in Fig.~\ref{beta_scan}, in this global treatment an artificial (i.e. not self-consistently modifying the background equilbrium) $\sim 40 \%$ increase in $\beta$ produces KBM instability in both the JET-ILW (92432) and JET-C (78697) pedestals for this wavenumber.  %Note that these scans did not self-consistently adjust the background equilibrium consistent with the $\beta$ variation and thus should not be considered a rigorous test of second stability.  %As discussed above, we find little evidence of KBM activity in the two pedestals of interest, but it cannot be ruled out.  In particular, MHD stability is potentially sensitive to additional effects such as the vacuum solution and kink term, both of which are neglected in our treatment.  %In any case, as described in Sec. x, the role of KBM would be limited to modifying the profile of the most weakly driven channel to keep the pressure profile near the KBM limit.  %Such dynamics are well-understood and have already found firm footing in the conventional pedestal paradigms of the community.  The purpose of this work is to elucidate the poorly understood role of non-MHD transport mechanisms in the pedestal, which we presently resume.        

In addition to the high $\beta$ KBM limit, a low $\beta$ threshold is also observed below which an electron drift wave becomes dominant.  This electron drift wave is excited by a parallel resonance and is suppressed by electromagnetic effects.  In the intermediate range, the slab ITG mode is very insensitive to $\beta$.  This suggests that, while certain pedestal modes are completely electrostatic in nature, it is necessary to include some level of electromagnetic effects in order to suppress spurious (i.e., only unstable at $\beta$ values far below the experimental values) electrostatic modes.

\section{Nonlinear Simulations} \label{nonlinear}

Having characterized the linear instabilities identified in the JET-C (78697) and JET-ILW (92432) pedestals, we turn now to the nonlinear simulations that are necessary to quantify how these instabilities are manifest in the transport.  % the transport levels produce by the instabilities described above, and (2) how the important parameter dependences discussed above translate into variations in transport.  
We employ a flux tube approximation for the small scale ETG turbulence and a global approach for the ion scale turbulence.  Although cross-scale interaction can not be ruled out, fully self-consistent multiscale simulations must await a large dedicated computer allocation.  Local neoclassical simulations are also carried out with \textsc{Gene}~\cite{doerk_dissertation}.  Detailed information about the simulations for these various modes of operation is provided in ~\ref{appendix}.  

As will be described below, simulations find close agreement (except near the separatrix) with carefully diagnosed~\cite{field_18} inter-ELM heat flux levels.  Such agreement is by no means a foregone conclusion and adds credence to the conclusions of this paper.  %Nonetheless, more rigorous validation remains to be done and will be the focus of future work.  
 Several elements are necessary to achieve such realistic transport levels.
 (1) Although the ITG turbulence described below is electrostatic and insensitive to $\beta$, some level of electromagnetic effects (i.e. finite $\beta$) is necessary to suppress the low $\beta$ electron drift waves that would otherwise dominate (see discussion surrounding Fig.~\ref{beta_scan}).  For other instabilities, like MTM and KBM, full electromagnetic effects (i.e., the precise experimental value of $\beta$) would be required.  (2) A global treatment is necessary to capture the limited domain of the steep gradient region.  This eliminates the large radial structures that dominate local pedestal flux tube simulations and produce unrealistically large transport levels.  The mechanism is closely related to the those described, e.g., in Ref.~\cite{mcmillan_10,goerler_11a}.  (3) A global treatment is additionally necessary to quantitatively capture the effects of the radial variation of geometry, magnetic shear, and profiles.  In our experience the limited radial domain (2) is an absolute necessity, whereas many semi-quantitative features can be captured with a local treatment that enforces a limited radial domain (with Dirchlet boundary conditions) corresponding to the pedestal width, as described in Refs.~\cite{hatch_16,hatch_17}.

\subsection{ETG Turbulence: Comparison of JET-ILW and JET-C} \label{ETG}

%As a small scale turbulence mechanism that is insusceptible to shear suppression, ETG has long been identified as a highly plausible pedestal transport mechanism.  
Due to its small spatial scales, ETG turbulence is amenable to nonlinear local flux tube simulations, which are described in this subsection.  ETG simulations are sensitive to not only temperature and density gradients, but also the temperature ratio and $Z_{eff}$, all of which are taken directly from the best available experimental estimates.  In agreement with earlier work~\cite{told_08,jenko_09,hatch_15,hatch_16}, the pedestal ETG turbulence described here is slab-like and isotropic, in contrast with the streamer-dominated core ETG turbulence.  

\begin{figure}[htb!]
 \centering
 \includegraphics{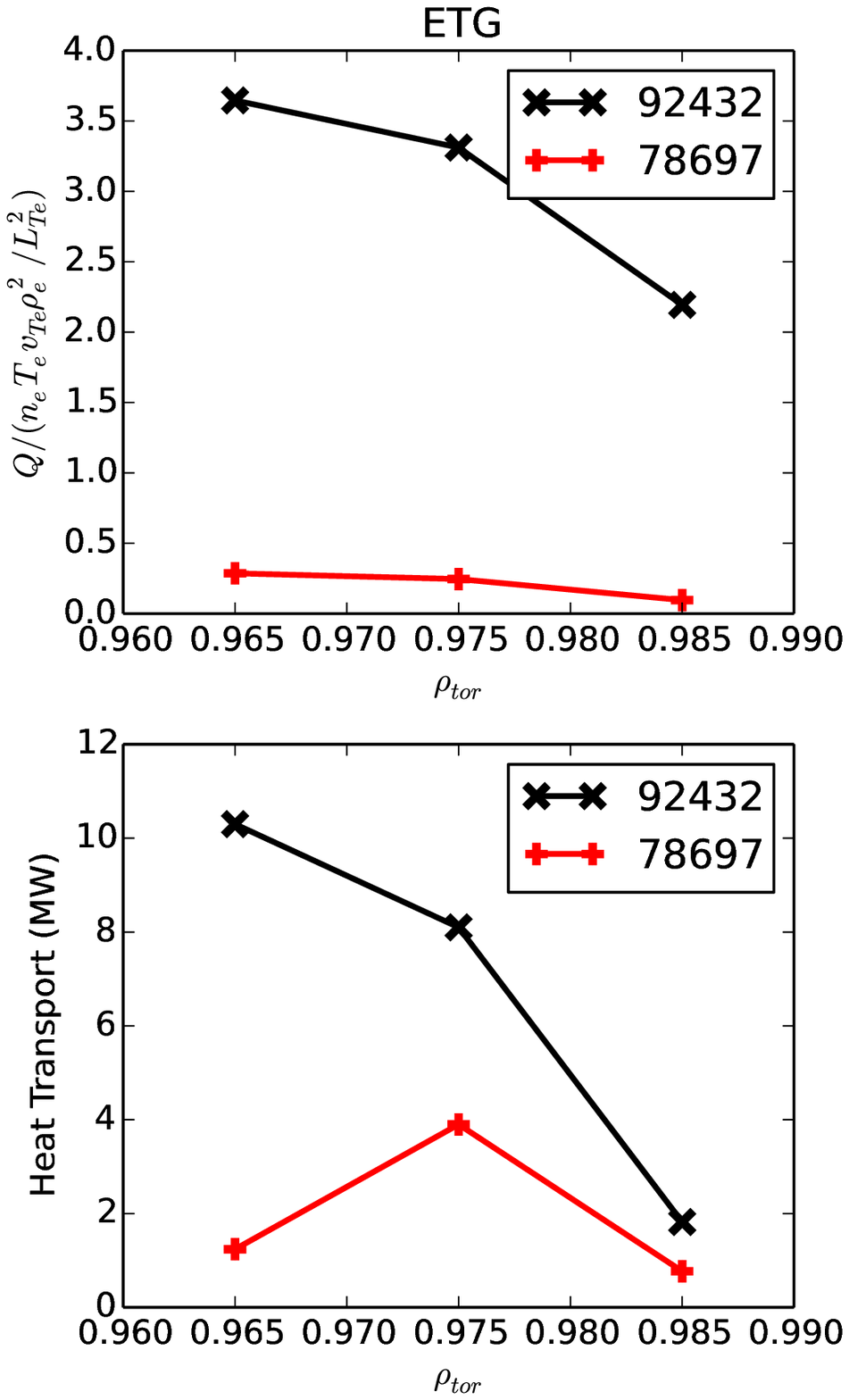}
 \caption{\label{ETG_plot}  Heat fluxes (top) from local flux tube single scale ETG simulations for JET-ILW (92432) (black) and JET-C (78697) (red) normalized to electron gyroBohm units (including the electron temperature gradient scale length).  As expected from the profiles (notably higher $\eta$), JET-ILW (92432) produces substantially higher heat fluxes, although the difference is decreased in the un-normalized quantities (bottom).  These simulations suggest that ETG produces experimentally relevant transport levels in both discharges (for reference the inter-ELM power is 11.6 MW and 5.7 MW for ILW and C, respectively).  }
\end{figure}

Consistent with the large difference in pedestal $\eta_e$, JET-ILW (92432) produces order of magnitude larger gyroBohm-normalized ETG heat fluxes than JET-C (78697), as shown in Fig.~\ref{ETG_plot} (top) for three points in the pedestal.  The electron gyroBohm normalization~\cite{hatch_15} using the electron thermal velocity, electron gyroradius, and electron temperature gradient scale lengths appears to be very appropriate, as it produces (roughly) order-unity fluxes.  As noted, for example, in Refs.~\cite{reshko_08,hatch_17,kotschenreuther_17}, ETG is sensitively dependent on $Z_{eff}$, which tends to be lower in JET-ILW than JET-C and would improve with increased $Z_{eff}$ associated with impurity seeding.

Although the normalized quantities differ by an order of magnitude, the raw heat transport is much more comparable between the two cases, as shown in Fig.~\ref{ETG_plot} (bottom), which shows the ETG heat transport in units of MW.  Notably, as a fraction of the total heat flux, the ETG transport is roughly equal for JET-ILW (92432) and JET-C (78697) (recall that the inter-ELM power losses are approximately twice as large for 92432---see Table~\ref{tab1}).  The fact that, despite widely varying gradients and gyroBohm fluxes, the two systems produce transport levels comparable to the experiment strongly suggests that ETG is an important transport mechanism in the JET pedestal. 

%The large difference between the gyroBohm normalized and un-normalized fluxes highlights the challenge inherent in achieving high pedestal temperatures: not only does it require steeper gradients (the SOL electron temperature cannot far-exceed $\sim 100 eV$ due to rapid parallel heat losses to the divertor), but also must overcome the natural gyroBohm scaling of the turbulence.  As will be discussed below, ITG turbulence plays a similar role, but also has unique parameter dependences distinct from ETG.  

%As seen from Fig.~\cite{gamma_ky} and further explored in the next subsection, the increased transport for JET-ILW is also observed at ion scales.     

%Nonetheless, basic considerations suggest that the JET-C ITG turbulence would be negligible.  As seen in Fig.~\ref{gamma_ky}, normalized JET-C ITG growth rates are approximately an order of magnitude smaller than the comparable JET-ILW growth rates.  However, due to the difference in gyroBohm heat flux, which is also approximately an order of magnitude, one might suspect comparable net transport levels similar to the scenario discussed above for ETG turbulence.  There are two additional factors that likely suffice to eliminate ITG transport as a major component in the JET-C pedestal: (1) the order of magnitude larger shear rates on JET-C (see Fig.~\ref{profiles} (e)), and (2) the fact that we have assumed $T_i = T_e$ for JET-C when in reality the ion temperature gradient is likely somewhat less steep than the electron temperature gradient.  %Note ratio of growth rates is comparable to ratio of QGB, suggesting that ion scale may play a comparable role in JET-C, but the low value of $\eta$ allows the achievement of high temperature.  

\subsection{ITG Turbulence in the JET-ILW Pedestal} \label{ITG}

In this section we describe the ion scale ITG turbulence, which may be the transport mechanism that is most distinctive to JET-ILW and responsible for its differences from standard pedestal regimes (JET-C and other experiments).  Although ITG turbulence is the major ion-scale instability in this JET-ILW discharge (92432), many of the dynamics may apply to other ion scale electrostatic modes like electron drift waves and TEM.  Within the framework described in Sec.~\ref{transport_paradigm}, such turbulence is likely robustly suppressed in both JET-C and other experiments (metal wall and otherwise).  For smaller experiments (i.e., higher $\rho_*$), ion scale transport is typically suppressed by the correspondingly stronger $E \times B$ shear rates.  Whereas for JET-C, such turbulence is likely suppressed due to some combination of reduced $\eta$, stronger shear rates (due to steep gradients), and stronger ion dilution from C impurities. 

%Nonetheless, the large difference in ITG growth rates (approximately and order of magnitude, as seen in Fig.~\ref{gamma_ky}) suggests that ion-scale electrostatic turbulence plays a minor role in JET-C.  

The ITG turbulence described below is similar to that investigated in detail in Refs.~\cite{hatch_17,kotschenreuther_17,hatch_18}: slab-like ITG turbulence that closely follows basic theoretical predictions for its response to shear flow~\cite{hatch_18}.  In some simulations, $\beta$ was reduced (up to $40 \%$) in order to avoid numerical instabilities.  Linear and nonlinear simulations verified that this reduction did not change the properties of the transport, since the slab ITG turbulence is insensitive to $\beta$ (see, e.g., Fig.~\ref{beta_scan}).  Simulations span from just inside the pedestal top to near the separatrix ($\rho_{tor} = 0.94-0.995$, closely corresponding to the range shown in Fig.~\ref{profiles}) with narrow regions at each end dedicated to a buffer zone where gradients are smoothly flattened and fluctuations are gradually set to zero (see ~\ref{appendix} for simulation details).  

\begin{figure}[htb!]
 \centering
 \includegraphics{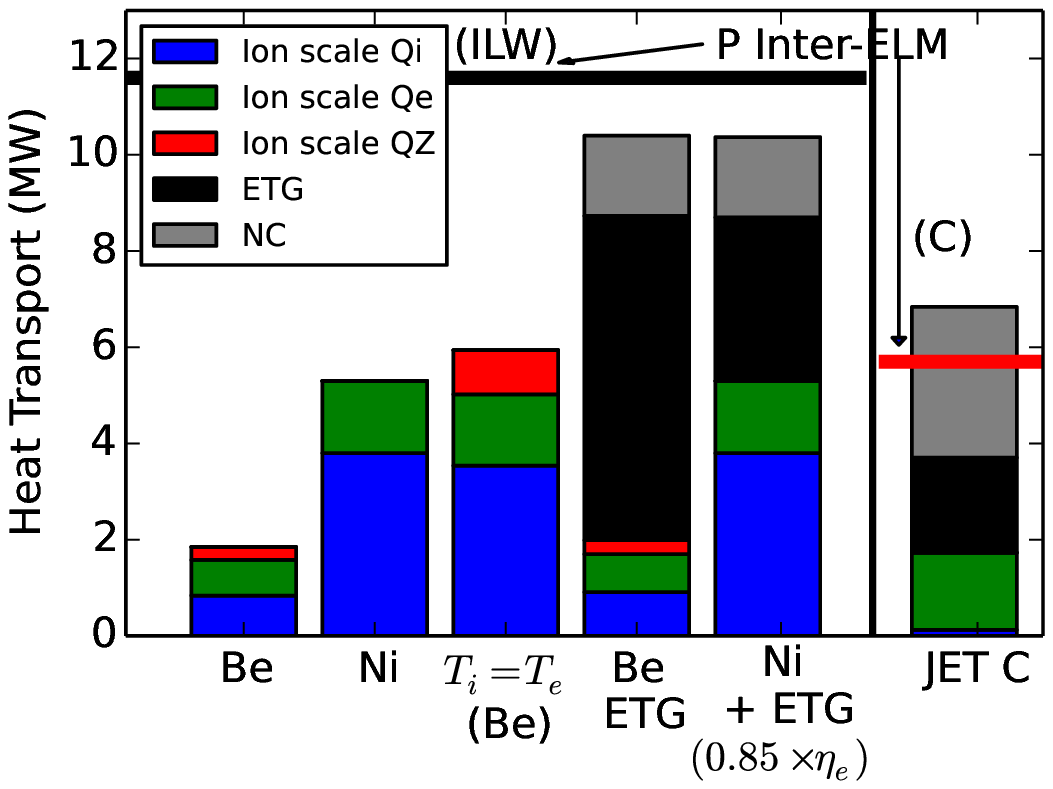}
 \caption{\label{transport}  Contribution to heat transport from global simulations of JET-ILW (92432) (first three bars), combined with ETG and neoclassical (bars four and five).  The first two columns probe sensitivity to ion dilution by using $Z_{eff}=1.8$ and two bounding assumptions for the impurity makeup: entirely beryllium (high dilution) and entirely nickel (low dilution).  The comparison of the third column ($T_i = T_e$) with the first column ($T_i$ from charge exchange, which estimates a $\sim 20\%$ decrease in $a/L_{Ti}$) tests sensitivity to the ion temperature gradient.  The fourth and fifth columns demonstrate two possible routes to reproducing the total power balance by combining transport from single scale ETG simulations and neoclassical simulations with global ion scale simulations.  The fourth column uses beryllium and the fifth column uses nickel along with a $15\%$ reduction in $\eta_e$ for the ETG simulations.  The last column shows transport from JET-C (78697) simulations (note the absence of turbulent ion heat flux).    }
\end{figure}

In order to realistically model the turbulence, the flow shear must be properly accounted for.  Direct measurements of the radial electric field $E_r$ and parallel flow $V_{||}$ were not available for this discharge.  Consequently, we estimate $E_r$ using the standard neoclassical formula~\cite{landreman_12},
\begin{equation}
	V_{||} = -\frac{R B_\phi}{Z e B} \left( \frac{1}{n_i} \frac{d P_i}{d \psi} + Z e \frac{d \Phi_0}{d \psi} - k_{||} \frac{B^2}{<B>^2}\frac{dT_i}{d \psi} \right )
\label{Er}
\end{equation}
where $V_{||}$ is the parallel flow, $R$ is major radius, $B_{\phi}$ is the poloidal magnetic field, $P_i$ is ion pressure, $\Phi_0$ is the electrostatic potential, and $\psi$ is the normalized poloidal magnetic flux (readers are referred to Ref.~\cite{landreman_12} for more detailed definitions, e.g., of $k_{||}$).  Since there is no measurement available, we use the approximation $V_{||}=0$, which is justified by the experimental observation~\cite{viezzer_13} that in the pedestal the dominant balance in Eq.~\ref{Er} is between the radial electric field and the gradients. Note that the Dopper shift in the pedestal is in the electron diamagnetic direction and thus opposite to that of the bulk plasma rotation in the core, so the inclusion of $V_{||}$ would be expected to slightly decrease the net Doppler shift.  We have calculated the effect of $V_{||}$ for similar JET-ILW discharges and find that it reduces the Doppler shift by a few ten percent (varying widely over the pedestal) and reduces the shear rate by $0-15\%$---i.e., it would produce a quantitative but not qualitative change in the turbulence.  

The shear rate used in {\sc Gene} is a flux function defined as 
\begin{equation}
\gamma_{GENE}  = \frac{\rho_{tor}}{q} \frac{d}{d \rho_{tor}}  \frac{E_r}{B_\theta R},
\label{shear_rate}
\end{equation}
where $\rho_{tor}$ is the square root of the normalized toroidal flux.  For global \textsc{Gene} simulations, the shear rate varies radially over the box and includes a region of zero shear as seen in Fig.~\ref{profiles} (e) (see Ref.~\cite{hatch_18} for a detailed discussion of pedestal $E \times B$ shear).  

Fig.~\ref{transport} shows the resulting heat transport for several variations of the simulations inputs.  The first two columns probe sensitivity to impurities by exploring the two limiting assumptions for impurity content: first, that the diagnosed value of $Z_{eff}$ is fully attributable to beryllium (i.e. maximum ion dilution), which is the material of the first wall, and second, that it is attributable to fully stripped nickel (i.e., minimal ion dilution).  Both treatments produce experimentally relevant transport levels.  The high sensitivity to ion dilution is demonstrated by the nearly three times increase in transport for the low dilution assumption.  The precise impurity mixture cannot be precisely determined experimentally, so that the most realistic transport estimate is expected to lie somewhere between these two bounds.  The third column in Fig.~\ref{transport} demonstrates sensitivity to the ion temperature gradient via a simulation using $T_i=T_e$, effectively increasing the ion temperature gradient by an average of $19\%$ over the pedestal (Ref.~\cite{haskey_PPCF_18} suggests that main ion temperature gradients may be larger than those inferred from impurity temperatures).  These simulations suggest that ion scale ITG turbulence (1) contributes experimentally relevant levels of transport, (2) is sensitive to impurity content, and (3) limits the ion temperature gradient and by extension the pedestal top temperature.  

\begin{figure}[htb!]
 \centering
 \includegraphics{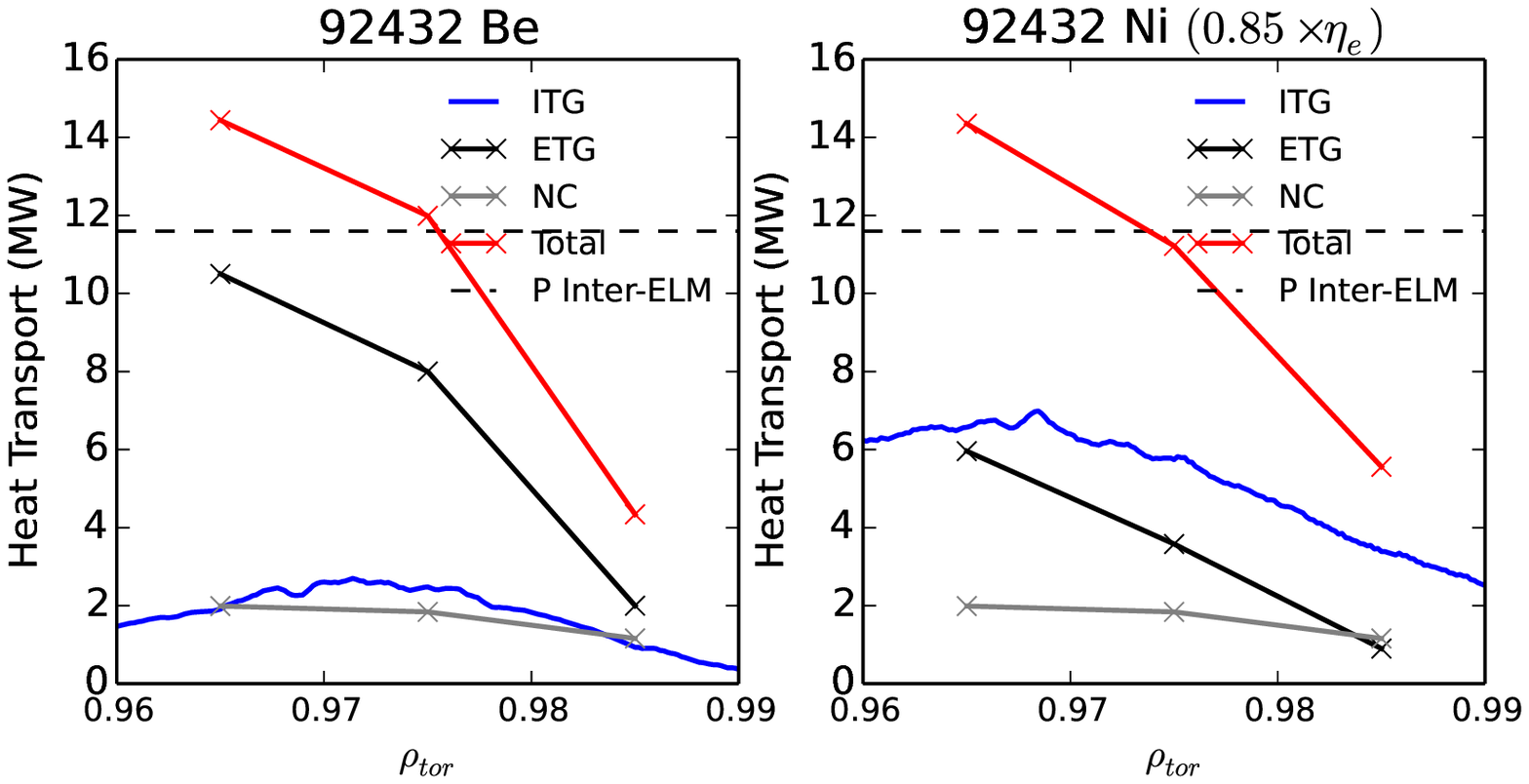}
 \caption{\label{radial_flux} Radial dependence of heat transport for 92432 using Be (left) and Ni (right).  The simulations match power balance closely at mid and upper pedestal but underpredict heat transport near the separatrix.  The discrepancy may be due to sensitivities to uncertain inputs (profile fits, impurity content, etc.), or alternatively an additional transport mechanism (e.g., blobs) that depends intrinsically on cross-separatrix dynamics. }
\end{figure}

The fourth (Be) and fifth (Ni) columns in Fig.~\ref{transport} show the sum of ion scale ITG transport, single scale ETG transport (with nominal profiles fourth and a 15 \% reduction in $\eta_e$ fifth), and neoclassical transport, demonstrating two plausible combinations that recover power balance.  The ETG and neoclassical transport are averaged over three points in the pedestal ($\rho_{tor} = 0.965, 0.975, 0.985$), and the global ion scale transport is taken to be the average from $\rho_{tor}=0.96-0.99$.  The radial dependence of the fluxes is shown in Fig.~\ref{radial_flux}.  For all transport channels, the fluxes decrease toward the outer region of the pedestal.  The total transport closely matches power balance in the mid and upper pedestal but underpredicts near the separatrix.  The discrepancy may be due to sensitivities to uncertain inputs (profile fits, impurity content, etc.), or alternatively an additional transport mechanism (e.g., blobs) that depends intrinsically on cross-separatrix dynamics.  The estimated inter-ELM power loss (11.6 MW) is also shown in Figs.~\ref{transport},\ref{radial_flux}.  Estimating this inter-ELM power loss,
\begin{equation}
P_{sep} = P_{abs}  - P_{rad}  - \frac{dW}{dt},
\label{power}	
\end{equation}
involves a careful accounting of the radiation from the confined plasma ($P_{rad}$), time-averaged ELM losses ($dW/dt$), and total absorbed heating power ($P_{abs}$)~\cite{field_18}. The time-averaged ELM loss power and the radiated power each account for about a third of the total loss power. Hence, the transport loss power between ELMs is substantially lower than the total power absorbed. 

%Notably, due to the ELM losses, this quantity is substantially below the naive estimate arrived at by simply subtracting the radiation from the heating power.

\subsection{Ion Scale Turbulence in JET-C} \label{JETC_iscale}

Global ion scale simulations for JET-C (78697) proved challenging due to two factors.  First, at realistic values of $\beta$, simulations either encountered numerical instabilities or (possibly related) proved extremely challenging due to the presence of MTM at low $k_y$.  Second, low $\beta$ simulations seeking to avoid these challenges were polluted by the spurious electrostatic electron drift waves that are suppressed by electromagnetic effects (described in Sec.~\ref{linear}, see Fig.~\ref{beta_scan}).  Stable simulations were achieved via a combined reduction of $\beta$ (by a factor of four) and the electron temperature gradient (by $30\%$).  These changes served the intended purpose of suppressing numerical instabilities, MTM, and the low $\beta$ electron drift wave while leaving the ITG growth rates unchanged.  The resulting transport, shown in the last column of Fig.~\ref{transport}, is composed almost entirely of electron heat flux, demonstrating the full suppression of ITG turbulence.  A substantial fraction of the heat flux is electromagnetic, suggesting that some MTM activity persists even at this low value of $\beta$.  Due to the substantial parameter modifications, which, in particular, affect the MTM instabilities, this simulation should not be considered a high-fidelity reproduction of the pedestal turbulence.  Nonetheless, it is sufficient to demonstrate the full suppression of ITG fluctuations in contrast with the JET-ILW (92432) pedestal described above.  This suppression was to be expected in light of the low ITG growth rates (Fig.~\ref{gamma_ky}) and high shear rates (Fig.~\ref{profiles} (e)) for 78697.    %The electron heat flux has a substantial electromagnetic component ($30 \%$) and may be the result of residual MTM instabilities.  %However, due to the large reduction in $\beta$ and $a/L_{Te}$, the simulation should not be considered a high-fidelity reproduction of such fluctuations.

\subsection{Stiffness and Shear Rate Scaling} \label{shear_dependence}

Typically ion heat flux in the pedestal is thought to be reduced to near-neoclassical levels~\cite{battaglia_14,viezzer_17}. This is consistent with the JET-C (78697) simulations described above.  In contrast, the JET-ILW (92432) simulations described above predict that neoclassical transport produces only between $30-70 \%$ of the ion heat flux (note that ETG remains a major, possibly dominant, transport mechanism).  Notably, consistent with the present study, a separate dedicated study of neoclassical transport for JET-ILW concludes that the neoclassical ion heat flux is low and unlikely to account for the experimental ion heat flux~\cite{giroud_18}.  

%The activity of ITG turbulence for JET-ILW has substantial implications, which are discussed in the next subsections.  

%Clearly, the transport levels simulated by \textsc{Gene} find close agreement with the experiment and exhibit strong sensitivity to details of the inputs.  This adds credence to the validity of the simulations and is encouraging for future efforts to invert the problem---i.e., to evolve pedestal profiles in response to the turbulent fluxes.

%\subsection{Stiffness and $\rho_*$ Scaling} \label{stiffness}

\begin{figure}[htb!]
 \centering
 \includegraphics{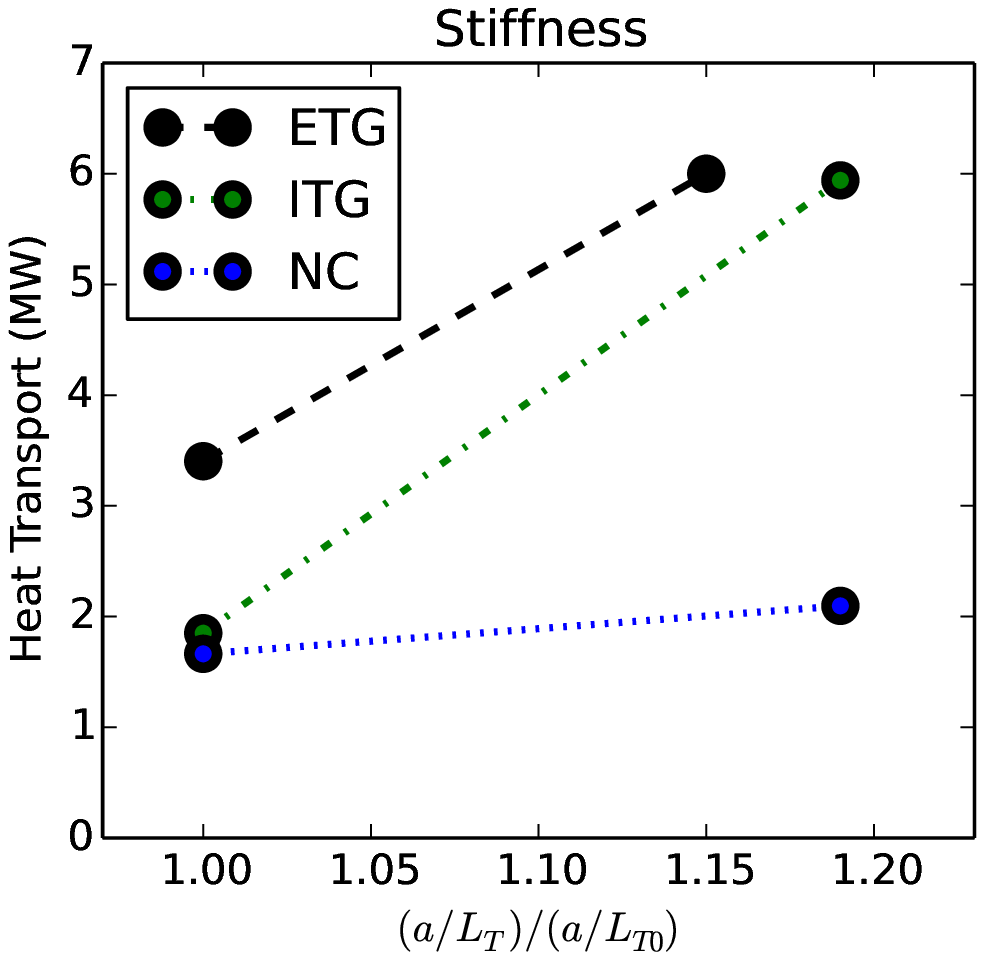}
 \caption{\label{stiffness_plot}  Stiffness (i.e. response of transport to driving gradients) of ETG, ITG, and neoclassical transport for 92432 (Be).  The ITG transport is extremely stiff in comparison with neoclassical, indicating the increased stiffness expected in pedestal regimes involving non-negligible ITG transport.  }
\end{figure}

Since neoclassical processes have generally been thought to account for pedestal ion heat flux, the addition of ITG turbulence would represent a qualitative change with respect to conventional pedestal regimes.  One implication is a substantial increase in stiffness (i.e. response of flux to gradients).  This is demonstrated in Fig.~\ref{stiffness_plot}, which shows the relative stiffness of ETG, ITG, and neoclassical heat transport.  In contrast with neoclassical transport, ETG and ITG turbulence exhibit a high degree of stiffness.  Note that this high gradient ITG simulation shown in Fig.~\ref{stiffness_plot} does not account for the increase in $E \times B$ shear, which decrease to some degree the observed stiffness.    %This again has implications for the increased demand for heating power found on JET-ILW.

The addition of ITG transport will also introduce a new $\rho_*$ sensitivity via the dependence of ITG transport on $E \times B$ shear rates.  Previous gyrokinetic studies have found that MTM, ETG, and neoclassical transport in the pedestal closely follow gyroBohm $\rho_*$ scaling ($Q \propto \rho_*^2$), which in turn closely follows empirical scaling laws~\cite{hatch_17,kotschenreuther_17}.  In these studies, pedestal ITG, on the other hand, was found to be strongly shear suppressed at high $\rho_*$ but exhibited a very unfavorable $\rho_*$ scaling, becoming non-negligible in the regimes of relevance for JET-ILW (low shear, high growth rates).  

\begin{figure}[htb!]
 \centering
 \includegraphics{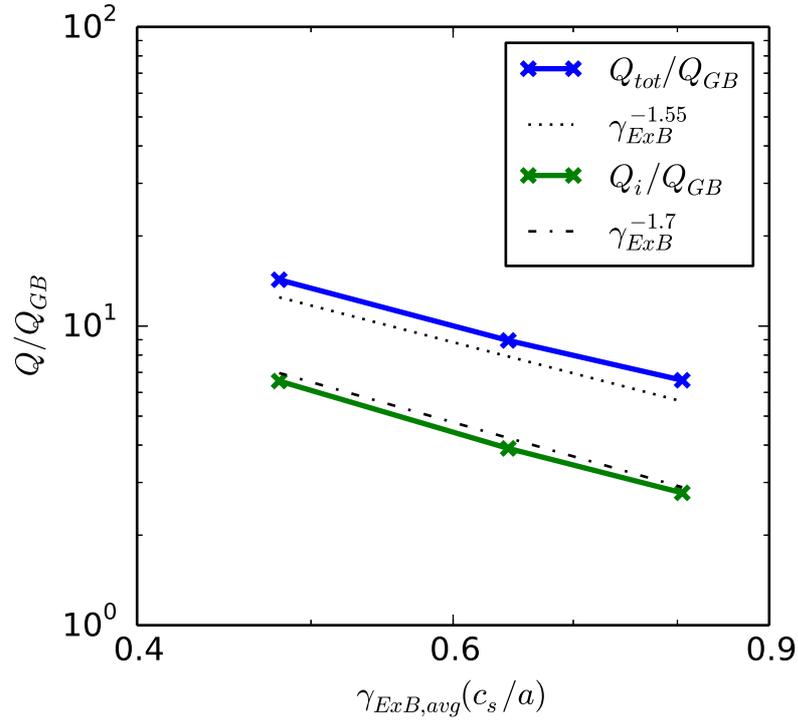}
 \caption{\label{shear_scan}  Heat flux from nonlinear global simulations of ITG turbulence for JET-ILW (92432) scanning $E \times B$ shear rate.  Both the ion heat flux (green) and total heat flux (blue) are shown.  The scaling is consistent with predictions from basic theory~\cite{hatch_18} and suggests high sensitivity to $\rho_*$.  Note that the total heat transport (including, e.g., ETG) would have a much weaker scaling. }
\end{figure}

Ref.~\cite{hatch_18} examines in detail this scaling of pedestal ITG transport on $\rho_*$ and $E \times B$ shear.  Since the pedestal ITG turbulence is slab like, many recent shear suppression models, which are oriented around the core-relevant toroidal picture, are less applicable and early slab decorrelation theories become highly relevant.  In particular, the decorrelation theory described in Ref.~\cite{zhang_92,zhang_93} finds excellent agreement with simulations and predicts the scaling $Q/Q_{GB} \propto \gamma_{E \times B}^{-2}$.  The ITG turbulence studied here also finds good agreement with this scaling as seen in Fig.~\ref{shear_scan}, which shows $Q/Q_{GB} \propto \gamma_{E \times B}^{-1.7}$ for the ion heat flux channel, and $Q/Q_{GB} \propto \gamma_{E \times B}^{-1.55}$ for the total ion scale transport.  We emphasize that this scaling applies only to the ITG component of the transport.  The total transport typically has large contributions from gyroBohm mechanisms like ETG, MTM, and neoclassical.  Such was the case described in Ref.~\cite{hatch_17}, where gyrokinetic simulations reproduced experimentally identified $\rho_*$ scaling~\cite{frassinetti_17} above a certain threshold in $\rho_*$.  

To put the $\rho_*$ scaling in context, the ITG transport would be up to $6$ times stronger for JET than for a system matching other dimensionless parameters and pedestal gradients on DIII-D or AUG (assuming a factor of three difference in $\rho_*$).  Likewise, within JET operational bounds, a factor of $\sim 2$ difference in ITG transport could be expected (assuming a factor of $\sim 1.5$ variation in $\rho_*$~\cite{nunes_13}).  The scaling could be even more severe if pedestal growth rates are enhanced at low $\rho_*$, e.g., by changes to the density profile in connection with more stringent operational constraints necessary to mitigate more energetic ELMs.  These dynamics are plausible contributing factors to both the more-severe confinement degradation (at least early in JET-ILW operation) due to metal walls on JET than AUG~\cite{beurskens_16} and the degradation on JET at low $\rho_*$.  Moreover, although no rigorous predictions can be made, these trends may portend a fundamental regime change for the ITER pedestal~\cite{kotschenreuther_17}.  Notably, such a change has also been described in Ref.~\cite{chang_17}, where the decrease in shear rate was found to result in streamers near the separatrix that broaden the SOL width.  Successful sustainment of the ITER pedestal may require pedestal profile manipulation targeted at decreasing pedestal growth rates and/or increasing pedestal shear rates.  Fortunately, due to the complex parameter dependences of drift-type microinstabilities, there exist many prospective routes toward such manipulation, which should be vigorously pursued. 

\begin{figure}[htb!]
 \centering
 \includegraphics{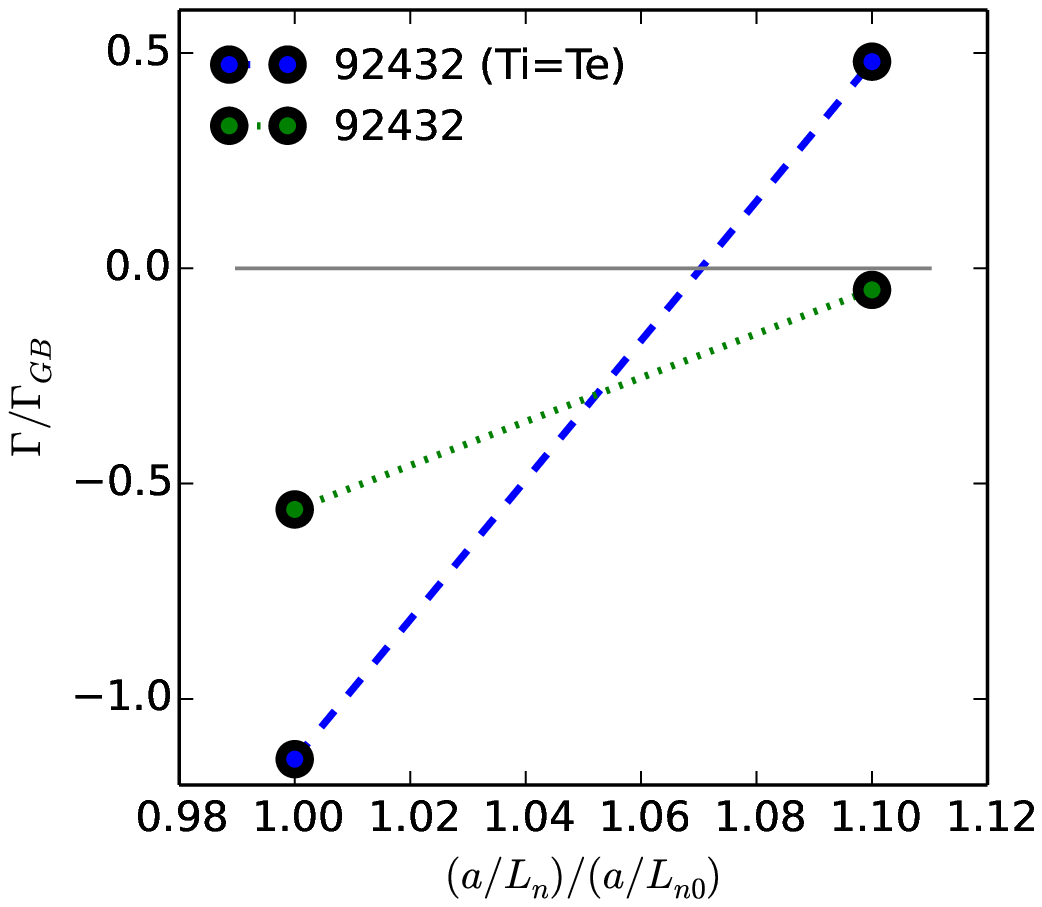}
 \caption{\label{pinch_plot}  Particle flux from nonlinear global simulations of two JET-ILW (92432) simulations.  The particle flux transitions from negative to zero or positive with a small increase in density gradient scale length, suggesting that the density pedestal reaches a state of balanced ITG pinch and diffusion. }
\end{figure}

\subsection{ITG Particle Pinch} \label{pinch}

%Experimental comparisons of particle transport are beyond the scope of this work due to the lack of information about pedestal particle sources.  As described in detail below, the ion scale JET-ILW ITG turbulence produces inward particle flux, which reverses direction with slight increases in density gradients.  In other words, ITG turbulence lies in a regime where pinch and diffusion are balanced and ion scale ITG turbulence should be considered a plausible candidate for producing a particle pinch and mediating the structure of the pedestal density profile.  

Particle fueling and transport in the pedestal remain large open questions, with some estimates predicting that an edge particle pinch or pellet fueling may be required to fuel ITER~\cite{kikushkin_03,garzotti_12}.  Here we discuss the potential role of ITG in pedestal particle transport.  The two best candidates for pedestal particle transport are (1) KBM, which produces only diffusion, and (2) ITG (or other ion scale electrostatic modes), which is versatile enough to produce any combination of pinch and diffusion depending on the details of the parameter regime~\cite{angioni_12}.  Interestingly, the ITG turbulence simulated in this work finds itself very near the point of balanced pinch and diffusion, as shown in Fig.~\ref{pinch_plot}, which shows the particle transport from global nonlinear simulations over density gradient scans for two JET-ILW (92432) simulations: the nominal profiles with Be (see first column in Fig.~\ref{transport}) and the case with $T_i=T_e$ (see third column in Fig.~\ref{transport}).  The scans shown in Fig.~\ref{pinch_plot} entail a $10\%$ increase in the density gradient scale length with a corresponding decrease in temperature gradients calculated to keep pressure fixed.  Notably, with this moderate change, the transport transitions from negative to positive (or nearly so) for all both cases.  This suggests that the ITG turbulence may be responsible for a particle pinch early in the ELM cycle and evolves to a state of balanced pinch and diffusion set by the properties of the ITG mode.  More thorough analysis of these dynamics would rely on detailed examination of inter-ELM profile evolution in connection with edge modeling to estimate particle sources, which will be the topic of future work.    

This ITG mechanism may be distinct to JET-ILW or may be operative in other pedestal scenarios.  Since the other transport mechanisms (ETG and MTM) produce negligible particle transport, KBM and/or ITG may operate at low fluctuation levels (negligible for heat flux) and still mediate the density profile.  

\section{Summary and Discussion} \label{summary}

This paper describes a comparison between the gyrokinetic instabilities and resulting transport produced in two representative JET-C and JET-ILW pedestals (shots 92432 and 78697, respectively).  The two discharges were selected to have high confinement at high current.  They have similar values for many relevant parameters ($\beta$, $I_p$, $B$, etc.) while retaining the distinguishing features of JET-C and JET-ILW, notably, decreased pedestal top temperature for JET-ILW.  A comparison of the profiles and heating power reveals a stark qualitative difference between the discharges: JET-ILW (92432) requires twice the heating power to sustain roughly half the temperature gradient of JET-C (78697).  This points to the heat transport as a central feature of the dynamics underlying the limitations on the JET-ILW pedestal.  This paper focuses on the relevant heat transport mechanisms, their important parameter dependences, and the interplay between these transport mechanisms and observed JET-ILW dynamics.    

This work builds on Ref.~\cite{hatch_17} and expands on those results by, among other things, (1) directly comparing JET-ILW with JET-C pedestal transport, (2) using an experimentally diagnosed ion temperature profile, (3) employing global electromagnetic nonlinear gyrokinetic simulations, (4) identifying direct connections between MTMs and magnetic fluctuations in JET-C, and (5) identifying an ITG particle pinch.  It reinforces the following emerging JET-ILW pedestal transport paradigm, which is proposed for further study.  ILW conditions modify the density pedestal in ways that preferentially decrease the pedestal density gradient.  This is attributable to some combination of direct effects of the metal wall on particle sources and the gas puffing necessary to mitigate W contamination.  The modification to the density profile increases $\eta_i$ and $\eta_e$, thereby producing more robust ITG and ETG instability which, in turn, limit the pedestal temperature and demand more heating power to achieve good pedestal performance.  The decreased density gradient also decreases the flow shear rate, doubly enhancing the ion scale transport.  The resulting decrease in pedestal temperature generally produces less favorable MHD stability and ultimately limits the pressure pedestal as well.  This paradigm points to understanding and manipulating SOL and pedestal density as the key to optimizing pedestal performance for JET-ILW.  

%Several ITG and/or ETG parameter dependences are consistent with additional JET-ILW observations, including the reduction of instability with impurities (via ion dilution for ITG and higher $Z_{eff}$ for ETG), and the more severe metal-wall constraints on JET as opposed to AUG (due to $\rho_*$ dependence of ITG turbulence).  

%This is the operating hypothesis of Refs.~\cite{hatch_17} and is further supported by the present study. 
%The H-mode edge is governed by three interconnected phenomena: (1) pedestal MHD stability, (2) divertor and SOL conditions, and (3) the residual transport in the edge transport barrier.  Of these three, transport has been perhaps the most poorly understood but, nonetheless, may provide the missing information for explaining JET-ILW.  %The turbulent transport determines the inter-ELM trajectory of density and temperature profile evolution, which, in turn, determines the operating point at which MHD stability limits are surpassed and an ELM is triggered.    
%Due to the well-established JET-ILW temperature limitation and the generally observation that the heat source far surpasses the particle source in the pedestal, a transport mechanism that preferentially produces heat transport is the most likely culprit.  

%JET-ILW is characterized by higher $\eta$, leading to much higher growth rates and correspondingly larger transport from drift-type instabilities ITG and ETG.  The higher values of $\eta$ are most likely attributable to the changes in SOL conditions due to the ILW that modify the density profile (see, e.g., Wolfrum).

Nonlinear simulations of ETG and ITG transport for JET-ILW (92432), in combination with neoclassical ion heat flux, are compatible with carefully diagnosed inter-ELM heat losses.  Global nonlinear simulations predict that ITG transport accounts for between $\sim17-47\%$ of the inter-ELM transport depending on assumptions regarding impurity content.  There are considerable uncertainties involved in diagnosing the ion temperature profile.  Despite these uncertainties, the ILW equilbrium studied here clearly lies in a regime that, in comparison to standard pedestal scenarios, favors ITG turbulence.  Consequently, these predictions of ITG turbulence should be taken seriously and be subject to continued examination via theory/simulation and experiment.  

ITG and ETG turbulence are sensitive to the density gradient, impurities and ion dilution, and ITG is additionally sensitive to flow shear.  Anomalous transport from these sources of temperature gradient driven turbulence could help explain several JET-ILW trends including: 
%The sensitivity of ITG turbulence to flow shear rates (proportional to $\rho_*$), temperature gradients, and ion dilution (i.e. impurity content) make it a highly plausible contributing factor to several JET-ILW trends, including: 
the strong limitation on accessible pedestal top temperatures; the observed degradation at high current / field (i.e. low $\rho_*$); and the increased confinement observed with impurity seeding.  Scans of $E \times B$ shear predict the scaling $Q_i/Q_{GB} \propto \rho_*^{-1.7}$, finding close agreement with the scaling predicted by fundamental theory~\cite{hatch_18} and suggesting that, if $a/L_{Ti}$ is fixed and above the ITG threshold, ITG turbulence will increase in impact as $\rho_*$ decreases.  Note that this scaling is only predicted for the ITG component of the transport and not the total transport, which also includes substantial contributions from ETG.  The presence of ITG also has potential implications for pedestal particle transport and pedestal density structure; nonlinear simulations scanning the density gradient find ITG to be in a regime of closely balanced pinch and diffusion, suggesting that it may be responsible for an inter-ELM particle pinch.  To the extent that the JET-ILW pedestal is characterized by high $\eta$ and low $E \times B$ shear, it lies in a unique parameter regime among present-day experiments that favors the excitation of ITG turbulence in the pedestal.

In contrast with JET-ILW (92432), our analysis suggests that JET-C (78697) is much more conventional in its composition of pedestal transport mechanisms, with heat transport dominated by a combination of ETG and MTM.  ETG simulations predict substantial levels of transport in the pedestal (roughly half the inter-ELM power loss).  MTMs are identified in global linear simulations and find close connections with washboard modes, which have long been observed in JET fluctuation data.  A detailed comparison between MTM and magnetic spectrograms will be presented in a future paper. This mix of ETG and MTM transport appears to be quite common in the pedestal, and is similar to the findings of Ref.~\cite{kotschenreuther_18}, which report similar analysis of two DIII-D discharges.  Moreover, the MTMs observed here are similar to those reported in a gyrokinetic analysis of another JET-ILW shot~\cite{hatch_16}.  In short, there is growing evidence that MTM is a common pedestal fluctuation across different operating scenarios on several machines. 

The results described here are not in direct conflict with MHD-centric models of the pedestal like EPED.  Such models may be consistent with the pre-ELM pressure pedestal structure and yet have limited predictive or explanatory power for scenarios like JET-ILW.  For example, EPED uses pedestal top density as an input to the model.  Consequently it is incapable of predicting or interpreting the observed JET-ILW changes in pedestal top density (note that a generalization of this framework has been developed for JET to get around such limitations~\cite{saarelma_18}).  Moreover, the question of the increased demand for heating power on JET-ILW is outside the scope of the EPED model.  The main intersection of this work with EPED is analysis of KBM stability.  In our analysis, KBM is found in local but not global simulations.  The discrepancy can be attributed to the fact that the local mode structure is too broad radially to fit inside the narrow radial domain of the pedestal and thus would not be manifest in a global simulation.  We note also, some remaining limitations in the global treatment, notably, the absence of the kink term and neglecting the vacuum solution beyond the separatrix.  A possible indication of MHD activity is the clamping of the pressure profile midway through the ELM cycle.  However, due to the underlying properties of MHD modes, such modes would be limited to constraining the pressure profile by preferential producing transport in the most-weakly driven channel (particles).  Consequently, KBM and other MHD modes can be eliminated as major heat transport mechanisms that constrain the JET-ILW pedestal temperature.

%This work adds additional data points on both the utility of gyrokinetics in the pedestal as well as remaining limitations.  The limitations stem from both computational challenges and challenges inherent in working with uncertain experimental data.  Remaining computational challenges include: modeling global low-n MTM turbulence; the question of ion-electron multiscale coupling; and sensitivities of gyrokinetic MHD stability calculations (including boundary conditions and the kink term).  Challenges related to experimental data include: strong sensitivity to highly uncertain experimental inputs; lack of experimental fluctuation data necessary for more rigorous validation; and difficulty in diagnosing transport fluxes and particle sources.  Despite these limitations, this work further substantiates gyrokinetics as a powerful tool for understanding pedestal transport.  Notably, heat transport predictions are firmly in the range of experimental expectations and agreement can be easily be found between experiment and simulations based on minor changes to input parameters, adding credence to the conclusions drawn in this work.  Such agreement is by no means a foregone conclusions as evidenced by the many examples in the literature where ultimate agreement could only be achieved by successive improvement of modeling capabilities.  The combination of realistic transport levels and high sensitivity to gradients suggests that inverting the problem---i.e., using gyrokinetic turbulence simulations to evolve profiles---is likely to predict realistic pedestal profiles.  

In summary, this work provides a framework for understanding the changes to pedestal transport that arise due to the ILW and limit the pedestal temperature.  This framework has the potential to inform pedestal optimization for JET-ILW and beyond and should be subject to ongoing examination by theory, computation, and experiment.  

\appendix

\section{Numerical setup of gyrokinetic simulations}
\label{appendix}

This work exploited several \textsc{Gene} modes of operation (local, global, linear, nonlinear, neoclassical), each with distinct numerical demands.  Numerical details of the simulations are described in this appendix.  All simulations were electromagnetic and employed Landau-Boltzmann collision operator with physical collision frequencies.   

\subsection{Global Simulations}

Global simulations use 320 radial grid points and span the domain shown in Fig.~\ref{profiles}: $\rho_{tor}=0.94-0.995$.  This corresponds to approximately 50 sound gyroradii (with the gyroradius calculated at the center of the radial domain).  Dirichlet boundary conditions were enforced at the radial boundaries and transition regions were implemented (15 \% on each side) over which gradients are smoothly set to zero and Krook damping smoothly ramps up to set fluctuations to zero at the boundary.  In nonlinear simulations, particle and heat sources (with coefficients of $0.05$ in the normalized time units) were employed to fix the gradients at their background values.     

Global simulations used 48 Fourier modes in $k_y$, with a minimum wavenumber $k_{y,min} = 0.041$ (i.e. $k_y$ ranges from 0.0 to 1.9).  Hyperdiffusion (6th order) was employed for $k_y$ modes in order to damp high $k_y$ ETG modes.  The hyperdiffusion was tuned to eliminate high $k_y$ electron modes while leaving the low $k_y$ ion scale turbulence unaffected.  Multiscale effects remain an open question in the pedestal and such simulations should be a high future priority to determine what, if any, affects of cross-scale coupling are important.  In the parallel $z$, parallel velocity $v_{||}$, and magnetic moment $\mu$ (i.e. squared perpendicular velocity), ($64,64,24$) grid points were used, respectively.  The parallel domain was from $-\pi$ to $\pi$ (poloidal angle), the parallel velocity domain was $-4$ to $4$ (normalized to $\sqrt{T_e/m_i}$), and the $\mu$ coordinate ranges from $0$ to $11$ (normalized to $T_e/B_0$).           

Numerical instabilities were encountered in some nonlinear global simulations, often after a substantial period of nonlinear saturation.  In such cases $\beta$ was slightly reduced and the simulations were extended successfully with no instability.  For the simulations of shot 92432, minor reductions succeded in extending the duration of the simulations and a $40 \%$ reduction appeared to eliminate the numerical instability.  These reductions in $\beta$ had no effect on turbulence levels since the ITG modes of interest are insensitive to $\beta$ in this regime (see Fig.~\ref{beta_scan}).  The simulations of shot 78697 were more challenging.  Minor reductions (a few ten percent) of $\beta$ were not sufficient to eliminate the numerical instability.  Larger reductions in $\beta$ ($>50\%$) eliminitated numerical instabilities but placed the system in the regime of spurious electron drift modes discussed in the context of Fig.~\ref{beta_scan}.          %Note, however, that a purely electrostatic treatment would not be accurate, as an electron drift mode is manifest in the electrostatic limit and stabilized by electromagnetic effects.  

Convergence tests in all coordinates were conducted by comparing simulation results through the initial saturation period and ensuring minimal changes with increases in resolution.

\subsection{Flux Tube ETG Simulations}

Single scale ETG simulations with adiabatic ions were conducted for both JET-C and JET-ILW using the local flux tube approximation.  $240, 36, 8$ grid points in the $z,v_{||},\mu$ coordinates, with a radial wavenumber grid resolving from $k_{x,min} \rho_s = 2.6$ to $k_{x,max} \rho_s = 164$ and a wavenumber grid in the $y$ coordinate resolving from $k_{y,min} \rho_s = 5$ to $k_{y,max} \rho_s=235$.  The parallel domain was from $-\pi$ to $\pi$ (poloidal angle), the parallel velocity domain was $-3$ to $3$, and the $\mu$ coordinate ranges from $0$ to $9$.  These simulations relied on convergence tests carried out in the course of earlier work~\cite{hatch_16}.           

\subsection{Local Linear Simulations}

Local linear simulations used $96, 48, 16$ grid points in the $z,v_{||},\mu$ coordinates along with 13 $k_x$ modes.  The parallel grid resolution was increased substantially ($\gtrsim 240$) for ETG simulations.  The parallel domain was from $-\pi$ to $\pi$ (poloidal angle), the parallel velocity domain was $-3$ to $3$, and the $\mu$ coordinate ranges from $0$ to $9$.           

\subsection{Neoclassical Simulations}

Local neoclassical simulations used $48, 96, 48$ grid points in the $z,v_{||},\mu$ coordinates.  The parallel domain was from $-\pi$ to $\pi$ (poloidal angle), the parallel velocity domain was $-3$ to $3$, and the $\mu$ coordinate ranges from $0$ to $9$.

{\em Acknowledgements.--} This research used resources of the National Energy Research Scientific Computing Center, a DOE Office of Science User Facility; the Texas Advanced Computing Center (TACC) at The University of Texas at Austin.  We acknowledge the CINECA award under the ISCRA initiative, for the availability of high performance computing resources and support.  We wish to acknowledge several members of the \textsc{Gene} development team, including F. S. Jenko, T. G/``orler, D. Told, and A. Ba\~n\'on Navarro, for useful support and discussion.   This work was supported by U.S. DOE Contract No. DE-FG02-04ER54742 and U.S. DOE Office of Fusion Energy Sciences Scientific Discovery through Advanced Computing (SciDAC) program under Award Number DE-SC0018429. This work has been carried out within the framework of the EUROfusion Consortium and has received funding from the Euratom research and training programme 2014-2018 under grant agreement No 633053. The views and opinions expressed herein do not necessarily reflect those of the European Commission.

%\begin{figure}[htb!]
% \centering
% \includegraphics{local_linear_x96.ps}
% \caption{\label{local_linear_x96}  Growth rates (top) and frequencies (bottom) from local linear simulations at the pedestal top for JET-ILW (black lines) and JET-C (red lines).  KBM, MTM, and electrostatic instabilities are observed.  The KBM have very broad radial mode structures that are inconsistent with the width of the pedestal.  The dashed lines, which find much closer agreement with global simulations, denote the instabilites that remain after filtering unphysically large radial modes.   }
%\end{figure}

\section{References}

\bibliography{my_refs}{}
\bibliographystyle{unsrt}

\end{document}